\documentclass[a4paper,fleqn]{cas-dc}

\usepackage[authoryear]{natbib}
\usepackage{lineno}

\usepackage{setspace}

\usepackage{ulem}

\def\tsc#1{\csdef{#1}{\textsc{\lowercase{#1}}\xspace}}
\tsc{WGM}
\tsc{QE}
\tsc{EP}
\tsc{PMS}
\tsc{BEC}
\tsc{DE}

\begin{document}
\let\WriteBookmarks\relax
\def\floatpagepagefraction{1}
\def\textpagefraction{.001}
\shorttitle{Mutual Phenomena between the Jovian moons}
\shortauthors{B. Morgado, R. Vieira-Martins, M. Assafin et~al.}

\title [mode = title]{The 2014-2015 Brazilian Mutual Phenomena campaign for the Jovian satellites and improved results for the 2009 events.}                      
\tnotemark[1]

\tnotetext[1]{Based in part on observations made at the Laborat\'orio Nacional de Astrof\'isica (LNA), Itajub\'a-MG, Brazil.}

\author[1,2,3]{B. Morgado}[                      orcid=0000-0003-0088-1808]
\cormark[1]
\ead{Morgado.fis@gmail.com}

\author[1,2,3]{R. Vieira-Martins}[]
\author[2,3]{M. Assafin}[]
\author[1,2,4]{A. Dias-Oliveira}[]
\author[5,6]{D. I. Machado}[]
\author[1,2]{J. I. B. Camargo}[]
\author[7]{M. Malacarne}[] 
\author[8]{R. Sfair}[]
\author[8]{O. C. Winter}[]
\author[9,1,2]{F. Braga-Ribas}[]
\author[1,2]{G. Benedetti-Rossi}[] 
\author[8]{L. A. Boldrin}[]
\author[8]{B. C. B. Camargo}[] 
\author[8]{H. S. Gaspar}[] 
\author[1,2,3,8]{A. R. Gomes-J\'unior}[]
\author[7]{J. O. Miranda}[] 
\author[8]{T. de Santana}[]
\author[6]{L. L. Trabuco}[]

\address[1]{Observat\'orio Nacional/MCTIC, R. General Jos\'e Cristino 77, Rio de Janeiro, RJ 20.921-400, Brazil}

\address[2]{Laborat\'orio Interinstitucional de e-Astronomia - LIneA and INCT do e-Universo, Rua Gal. Jos\'e Cristino 77, Rio de Janeiro, RJ 20921-400, Brazil}

\address[3]{Observat\'orio do Valongo/UFRJ, Ladeira Pedro Antonio 43, Rio de Janeiro, RJ 20080-090, Brazil}

\address[4]{Escola SESC de Ensino M\'edio, Avenida Ayrton Senna, 5677, Rio de Janeiro - RJ, 22775-004, Brazil}

\address[5]{Universidade Estadual do Oeste do Paran\'a (Unioeste), Avenida Tarqu\'inio Joslin dos Santos 1300, Foz do Igua\c{c}u, PR 85870-650, Brazil}

\address[6]{Polo Astron\^omico Casimiro Montenegro Filho/FPTI-BR, Avenida Tancredo Neves 6731, Foz do Igua\c{c}u, PR 85867-900, Brazil}

\address[7]{Universidade Federal do Esp\'irito Santo, Av. Fernando Ferrari 514, Vit\'oria, ES 29075-910, Brasil}

\address[8]{UNESP - S\~ao Paulo State University, Grupo de Din\^amica Orbital e Planetologia, CEP 12516-410, Guaratinguet\'a, SP 12516-410, Brazil}

\address[9]{Federal University of Technology - Paran\'a (UTFPR/DAFIS), Av. Sete de Setembro, 3165, Curitiba, PR 80230-901, Brazil}

\cortext[cor1]{Corresponding author}

\begin{abstract}
Progress in astrometry and orbital modelling of planetary moons in the last decade enabled better determinations of their orbits. These studies need accurate positions spread over extended periods. We present the results of the 2014-2015 Brazilian campaign for 40 mutual events from 47 observed light curves by the Galilean satellites plus one eclipse of Amalthea by Ganymede. We also reanalysed and updated results for 25 mutual events observed in the 2009 campaign. 

All telescopes were equipped with narrow-band filters centred at 889 nm with a width of 15 nm to eliminate the scattered light from Jupiter. The albedos' ratio was determined using images before and after each event. We simulated images of moons, umbra, and penumbra in the sky plane, and integrated their fluxes to compute albedos, simulate light curves and fit them to the observed ones using a chi-square fitting procedure. For that, we used the complete version of the Oren-Nayer reflectance model. The relative satellite positions mean uncertainty was 11.2 mas ($\sim$35 km) and 10.1 mas ($\sim$31 km) for the 2014-2015 and 2009 campaigns respectively. The simulated and observed \textsc{ascii} light curve files are freely available in electronic form at the \textit{Natural Satellites DataBase} (NSDB).

The 40/25 mutual events from our 2014-2015/2009 campaigns represent a significant contribution of 17\%/15\% in comparison with the PHEMU campaigns lead by the IMCCE. Besides that, our result for the eclipse of Amalthea is only the 4$^{th}$ such measurement ever published after the three ones observed by the 2014-2015 international PHEMU campaign. Our results are suitable for new orbital/ephemeris determinations for the Galilean moons and Amalthea.

\end{abstract}




\maketitle

\section{Introduction} \label{intro}

Mutual phenomena between natural satellites -- occultations and eclipses -- have been successfully used to improve the orbital studies of these moons. For the Galilean satellites, they have been systematically observed since 1976 \citep{Aksnes1976}. These phenomena occur as the Earth and the Sun cross the orbital plane of the satellites. For Jupiter, they happen every six years.

The photometry of these events offers a reliable source of very precise relative positions between two satellites. They often achieve uncertainties bellow 5 mas ($\sim$ 15 km) \citep{Emelyanov2009,Dias-Oliveira2013,Arlot2014,Saquet2018}. These relative positions can constrain the orbital studies of these moons and give us hints about their structure and formation processes \citep{Lainey2004a,Lainey2004b,Lainey2009,Lainey2017}.

The uncertainty of the positions obtained from mutual phenomena is usually smaller than the ones obtained by other methods. For instance, classical CCD astrometry achieves uncertainties around 100 mas ($\sim$ 300 km) \citep{Kiseleva2008}. For satellite-pair distances, the uncertainties are at the 30 mas level ($\sim$ 90 km) \citep{Peng2012}. Mutual approximations, based in the same geometrical configuration of mutual occultations, achieve uncertainties at the 10 mas level ($\sim$ 30 km) \citep{Morgado2016,Morgado2019}. 

In this paper, we present results for 47 light curves, 31 occultations and 16 eclipses, representing 40 mutual events between the Galilean moons observed by three stations in Brazil, during the 2014-2015 mutual phenomena campaign. We also present one event, an eclipse involving the inner satellite Amalthea (J5). We also used our improved methods to re-analyse 25 light curves, 13 occultations and 12 eclipses, of 25 mutual phenomena observed by our group during the 2009 mutual phenomena campaign. We compared the new results with those by \cite{Dias-Oliveira2013}, \cite{Arlot2014} and \cite{Morgado2016}.

In Section \ref{Obs} we detail the observational campaigns. In Section \ref{redu} we present the photometry used to produce the observed light curves and describe the new, improved light-curve fitting procedures developed and used in this work. Section \ref{compar_2009} contains new results from the re-analysis of 25 mutual events observed in 2009 and the comparison with the older results. In Section \ref{res2015}, we present the results for the 47 light curves involving 40 mutual events observed from Brazil during the 2014-2015 campaign. In Section \ref{resAmalthea}, we present the result for the eclipse involving Amalthea. Our conclusions are set on Section \ref{conclusao}.

\section{Mutual phenomena campaign details} \label{Obs}

Every six years, during Jupiter equinox, we can observe mutual occultations and eclipses between Jupiter's regular satellites. The results presented here come from the collaboration between five Brazilian institutes. The prediction of these events was provided by the \textit{Institut de M\'ecanique C\'eleste et de Calcul des Eph\'em\'erides} (IMCCE)\footnote{Website: \url{http://nsdb.imcce.fr/multisat/nsszph515he.htm}} \citep{Arlot2014,Arlot2014b}. 

The 2009 mutual phenomena campaign was the first large attempt of the kind carried out in Brazil for the Galilean moons. Observations and instruments are described in detail in \citet{Dias-Oliveira2013}. We re-analysed 25 mutual events encompassing 25 light curves, 13 from occultations and 12 from eclipses, and obtained new results for this campaign. Discrepancies between the results obtained by \cite{Dias-Oliveira2013} and by \cite{Arlot2014} motivated this re-analysis, see more details in Section \ref{compar_2009}.

The last Brazilian mutual phenomena campaign of 2014-2015 obtained data from three telescopes spread on the South and South-East of Brazil, with apertures ranging between 28 and 60 cm. We obtained 47 light curves, 31 for occultations and 16 for eclipses, from 40 events observed between November 2014 and June 2015. In Table \ref{tb:observers} we present the stations, observers and instrumental details of each station. It also contains the number of light curves obtained by each observatory. Moreover, we added the Minor Planet Center (MPC) observatory code of the station (XXX for the station without a code).     

In Table \ref{tb:obs_info}, we list the observational details of each observed event. It contains the date of the event and the satellites' pairs in the form "$S_{1} o S_{2}$" for occultation and "$S_{1} e S_{2}$" for eclipses, where 1 stands for Io, 2 for Europa, 3 for Ganymede and 4 for Callisto. We furnish the sites involved in each observation (using the alias defined in Table \ref{tb:observers}). For each event, we give the solar phase angle ($i$) and the zenith distance ($z$), both in degrees. In the last column, we list the instrumental albedos' ratio of the involved satellites (and its uncertainty), determined by using images before and after the event. This albedos' ratio is only needed for occultations.

\begin{figure}
\centering
\includegraphics[width=0.47\textwidth]{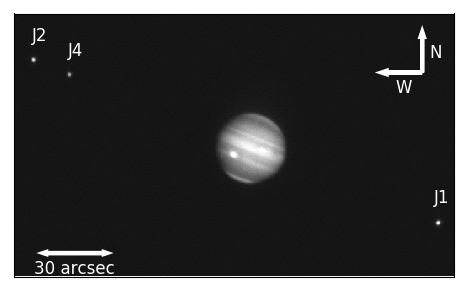}
\caption{Image of Jupiter, Io (J1), Europa (J2) and Callisto (J4) obtained with the 0.6 m diameter Zeiss telescope of the \textit{Observat\'orio Pico dos Dias}, equipped with a methane filter on 2014 November 19. The planet and the satellites present about the same brightness due to the use of the narrow-band filter, centred at $\lambda_0$ = 889 nm with 15 nm width.}
\label{Fig:jup_obs}
\end{figure}

In all observations we used a narrow band filter centred at 889 nm with a width of 15 nm. This bandpass is in the methane absorption region of the spectrum. We chose this filter because in this spectral region, Jupiter's albedo drops to 0.05 due to the absorption in the upper atmosphere \citep{Karkoschka1994,Karkoschka1998}. Figure \ref{Fig:jup_obs} shows an example of an image obtained with the 0.6 m Zeiss telescope from OPD. This filter has been successfully used in the 2009 mutual phenomena campaign \citep{Dias-Oliveira2013} and the mutual approximation campaigns started in 2014 \citep{Morgado2016,Morgado2019}.

\begin{table*}
\begin{center}
\caption{2014-2015 mutual phenomena campaign observer list.}
\begin{tabular}{lcccc}
\hline
\hline
City/ Country  & Longitude & Observers & Telescope & N$^{o}$ of positive \\
Station alias & Latitude &&CCD&detections \\
MPC code & Height && &         \\
\hline
\hline
Itajub\'a/ MG, Brazil     &45$^o$ 34' 57" W & B. Morgado         & 60 cm   & \textbf{20}  \\
OPD             &22$^o$ 32' 07" S  & H. S. Gaspar        & Andor/IKon-L    &    \\
874             &1.864 km          & R. Sfair             &           &    \\ 
                &                  & B. C. B. Camargo     &         &    \\
                &                  & T. Santana           &         &    \\
                &                  & L. A. Boldrin        &         &    \\
                &                  & M. E. Huaman         &         &    \\
                &                  & G. Benedetti-Rossi   &         &    \\
                &                  & A. R. Gomes-J\'unior &         &    \\
\hline
Foz do Igua\c{c}u/ PR, Brazil&54$^o$ 35' 37" W  & D. I. Machado      & 28 cm   & \textbf{22} \\
FOZ             &25$^o$ 26' 05" S  & L. L. Trabuco      & SBIG/ST-7X-ME    &    \\
X57                &0.184 km          &                    &         &    \\
\hline
Vit\'oria/ ES, Brazil      &40$^o$ 19' 00" W  & M. Malacarne       & 35 cm   & \textbf{5} \\
GOA             &20$^o$ 17' 52" S  & J. O. Miranda      & SBIG/ST-8X-ME    &    \\
XXX                &0.026 km          &                    &         &    \\
\hline
\hline
\label{tb:observers}
\end{tabular}
\end{center}
\end{table*}

\begin{table}
\begin{center}
\caption{Mutual events and observation conditions.}
\begin{tabular}{ccccccc}
\hline
\hline
Date  & Event & Obs. &  i  & z & Ratio of\\
yy-mm-dd  & $S_1$x$S_2$ & & ($^o$) & ($^o$) & Albedo     \\
\hline
\hline
14-11-02 & 4o1 & OPD & 10.60 & 73.49 & 3.56  $\pm$  0.03 \\ 
14-11-19 & 4o2 & OPD & 10.69 & 43.44 & 3.59  $\pm$  0.02 \\ 
14-12-20 & 2o1 & FOZ &  8.63 & 50.72 & 0.97  $\pm$  0.03 \\ 
14-12-21 & 4e1 & FOZ &  8.52 & 77.19 & -- \\ 
14-12-21 & 3o1 & FOZ &  8.52 & 65.78 & 1.65  $\pm$  0.19 \\ 
14-12-24 & 2e3 & FOZ &  8.13 & 40.85 & -- \\ 
15-01-21 & 2e1 & FOZ &  3.45 & 46.09 & -- \\ 
15-02-02 & 3o2 & FOZ &  0.95 & 67.15 & 1.53  $\pm$  0.09 \\ 
15-02-22 & 2o1 & OPD &  3.20 & 41.50 & 0.99  $\pm$  0.04 \\ 
15-02-22 & 2e1 & OPD &  3.20 & 43.44 & -- \\ 
15-03-01 & 2o1 & FOZ &  4.57 & 57.20 & 1.01  $\pm$  0.06 \\ 
15-03-01 & 2e1 & FOZ &  4.58 & 65.89 & -- \\ 
15-03-03 & 3o1 & OPD &  4.94 & 62.37 & 1.56  $\pm$  0.03 \\ 
15-03-03 & 3o1 & FOZ &  4.94 & 58.30 & 1.56  $\pm$  0.03 \\ 
15-03-06 & 1e2 & OPD &  5.46 & 43.08 & -- \\ 
15-03-09 & 3e2 & OPD &  6.14 & 44.04 & -- \\ 
15-03-13 & 1e2 & FOZ &  6.79 & 48.89 & -- \\ 
15-03-13 & 1e3 & FOZ &  6.66 & 60.76 & -- \\ 
15-03-16 & 4o2 & FOZ &  7.11 & 48.84 & 3.58  $\pm$  0.03 \\ 
15-03-17 & 3e2 & FOZ &  7.27 & 58.36 & -- \\ 
15-03-18 & 2e1 & GOA &  7.54 & 41.46 & -- \\ 
15-03-24 & 3o4 & OPD &  8.22 & 45.53 & 0.41  $\pm$  0.03 \\ 
15-03-24 & 3o4 & FOZ &  8.22 & 46.49 & 0.41  $\pm$  0.03 \\ 
15-03-25 & 2o1 & FOZ &  8.47 & 46.28 & 0.99  $\pm$  0.04 \\ 
15-03-25 & 2o1 & GOA &  8.47 & 42.75 & 0.99  $\pm$  0.04 \\ 
15-03-26 & 2e1 & OPD &  8.48 & 52.32 & -- \\ 
15-04-02 & 2o1 & OPD &  9.25 & 62.96 & 1.01  $\pm$  0.07 \\ 
15-04-02 & 2o1 & FOZ &  9.25 & 59.06 & 1.01  $\pm$  0.07 \\ 
15-04-02 & 2e1 & OPD &  9.26 & 82.45 & -- \\ 
15-04-03 & 1o3 & FOZ &  9.43 & 46.38 & 0.75  $\pm$  0.13 \\ 
15-04-06 & 1e2 & FOZ &  9.70 & 46.72 & -- \\ 
15-04-12 & 2e3 & FOZ & 10.09 & 65.99 & -- \\ 
15-04-14 & 1e2 & OPD & 10.22 & 69.43 & -- \\ 
15-04-17 & 4o1 & OPD & 10.43 & 54.20 & 3.90  $\pm$  0.08 \\ 
15-04-17 & 4o1 & GOA & 10.43 & 55.82 & 3.90  $\pm$  0.08 \\ 
15-04-18 & 4o3 & OPD & 10.48 & 44.23 & 2.32  $\pm$  0.05 \\ 
15-04-18 & 1o3 & OPD & 10.44 & 72.62 & 0.64  $\pm$  0.11 \\ 
15-04-25 & 1o3 & OPD & 10.73 & 58.34 & 0.69  $\pm$  0.07 \\ 
15-04-25 & 1o3 & FOZ & 10.73 & 55.16 & 0.63  $\pm$  0.07 \\ 
15-04-26 & 2o1 & OPD & 10.75 & 43.56 & 1.04  $\pm$  0.15 \\ 
15-04-29 & 3o1 & OPD & 10.80 & 67.89 & 1.60  $\pm$  0.06 \\ 
15-04-29 & 3o1 & GOA & 10.80 & 70.74 & 1.60  $\pm$  0.06 \\ 
15-05-03 & 2o1 & OPD & 10.85 & 61.83 & 0.96  $\pm$  0.07 \\ 
15-05-05 & 3o2 & FOZ & 10.85 & 47.24 & 1.61  $\pm$  0.06 \\ 
15-05-13 & 3o2 & OPD & 10.77 & 86.09 & 1.48  $\pm$  0.07 \\ 
15-06-04 & 2o1 & FOZ &  9.72 & 56.26 & 1.01  $\pm$  0.05 \\ 
15-06-18 & 3o1 & GOA &  8.59 & 59.64 & 1.58  $\pm$  0.04 \\ 
\hline
15-03-02 & 3e5 & 1.60 & 3.17 & 50.90 & -- \\ 
\hline
\hline
\label{tb:obs_info}
\end{tabular}
\end{center}
\textbf{\textit{Note}: The solar phase angle, zenith distance and ratio of albedo in the sense $S_2/S_1$ for each event.}
\end{table}

\section{Light curve analysis} \label{redu}

In mutual phenomena, one can determine relative positions between the satellites through the analysis of the events' light curves. In our procedure, we simulate theoretical light curves and use them to fit the observed ones.

The parameters of interest are: (i) the impact parameter ($s_{0}$), the smallest apparent angular distance in the sky plane between both satellite's centres in the case of occultations or between the eclipsed satellite centre and the centre of the eclipsing shadow in the sky plane for eclipses, both cases in a topocentric frame; (ii) the central instant ($t_{0}$), the instant of time that this smallest distance occurs; and (iii) the apparent relative velocity ($v_{0}$) between both satellites in the sky plane. 

In the supplementary material, we also provide the inter-satellite tangential coordinates ($X$, $Y$). For occultations, these coordinates between both satellites’ centres are in a topocentric frame. For mutual eclipses, these coordinates are in a topocentric frame and the mean difference between the eclipsed satellite centre and the centre of the eclipsing satellite’s shadow in the sky plane.

\subsection{Obtaining the observed light-curves} \label{photometry}

Firstly all images were corrected by Bias and Flat-Field using standard procedures with the Image Reduction and Analysis Facility (\textsc{iraf}) \citep{Butcher1981}. Then, we determined the light flux of the targets in the images by differential aperture photometry using the \textsc{praia} package \citep{Assafin2011}. During an occultation, both satellites are measured together in the same aperture, and a third satellite is used as calibrator. In the case of eclipses, the eclipsed satellite is measured alone in the aperture and the eclipsing satellite (or any other) is used as calibrator. The light curve is then normalised by a polynomial fit so that the flux ratio outside the flux drop gets equal to 1.0, and the flux drop can be adequately evaluated.

During the photometry of mutual events, one must take care with the possibility of a parasitic flux, as pointed out by \cite{Emelianov2017} and \cite{Arlot2017}. The origin of this flux is likely to be the background (mostly Jupiter's scattered light) or the CCD detector. In our case, we attenuate this parasitic flux with the Methane filter and a rigorous calibration process. Tests showed that the parasitic flux in our images is one order of magnitude below the noise of our observations.

\subsection{Simulating light curves}\label{simulation}

The procedures utilised here follow the same principles outlined in \cite{Assafin2009} and \cite{Dias-Oliveira2013}. However, improvements were made in almost every step, as explained in sections \ref{simulation_occ}, \ref{simulation_ecl} and \ref{fitting}.

The parameters needed in the modelling of a mutual occultation or eclipse can be separated into two complementary types. The first refers to physical characteristics: sizes and shapes of each satellite, and the satellites' albedo for occultations. The second type relates to dynamics, and are the parameters of interest: $s_{0}$, $t_{0}$ and $v_{0}$.

Physical parameters such as radius and shape are known from space probes' data. Albedos are determined from auxiliary observations made before and after events with the same instrument setup. The apparent separation in the sky plane between both satellites can be written as a function of the time and the dynamical parameters using equation \eqref{Eq:param_occ} \citep{Assafin2009}.

\begin{equation}
    s(t) = \sqrt{s_{0}^{2} + v_{0}^{2}(t - t_{0})^{2}} \label{Eq:param_occ}
\end{equation}

We also need a reflectance model to take into account the phase effect and how the surface of the satellite will reflect the sunlight. For eclipses, we further need some information about the Sun, such as its radius and a model to consider the Sun's limb darkening.

We could do simulations with triaxial bodies with varying albedo, but not for practical purposes, due to photometry limitations. Thus, satellites are considered as spheres with a known appropriate radius, and the albedo is uniform along the surface. The relative velocity between both satellites is constant during the mutual event (a few minutes only). For notation, the occulting/eclipsing is denoted \textit{Sat}$_{1}$ and \textit{Sat}$_{2}$ is the occulted/eclipsed one.

In the simulations of occultations and eclipses, we used the same geometric relations described in detail in \cite{Dias-Oliveira2013}.

\subsubsection{Occultation}\label{simulation_occ}

The first step in simulating a light curve of a mutual occultation is the production of a 2D satellite apparent profile, simulating how the body reflects the sunlight as seen by an observer on Earth. As pointed out by \cite{Vasundhara2017}, it is essential to use a realistic intensity distribution for the satellite.

However, this approach demands previous knowledge about the satellite surface (albedo maps) that can change with time or even for different effective wavelengths, it is important to highlight that these maps are not know for the wavelength of our observations ($\lambda_0$ = 889 nm). The same applies to the Hapke scattering law \citep{Hapke1981a,Hapke1981b,Hapke1984,Hapke1986,Hapke2002,Hapke2008,Hapke2012} used by \cite{Emelyanov2009}, \cite{Arlot2014} and \cite{Saquet2018}, which requires unknown parameters in the wavelength band of our observations.

We successfully solved these issues by adopting a generalisation of Lambert scattering law given by \citet{Oren1994}. The Oren-Nayer model takes into account the direction of radiance and the roughness of the surface in a natural way, so that the reflectance depends only of the albedo and in one more parameter that very smoothly tunes a wide range of surface roughness, and most importantly, regardless of the wavelength. This model realistically reproduces the illumination of an object in modern computer graphic scenes for movies and for the full Moon. \citep{Oren1994}. In \cite{Dias-Oliveira2013} a simplified version of the model was used. Here, we implemented the complete version in \citet{Oren1994}, taking into account the direct illumination and all inter-reflection components of the radiance.

The albedo ratio between the satellites is determined independently, by using observations right before and after the mutual event with the same instrument setup. In \cite{Dias-Oliveira2013}, analytic expressions involving the terminator were used to take solar phase angle effects into account in the evaluation of the flux measurements of albedo observations. Here, following a more rigorous approach, we also simulated the 2D profiles of the satellites for these observations, to better determine the effective area and reflectance of the satellites to compute more accurate albedos.

In fact, first we measured the light fluxes between both satellites ($F_1$, $F_2$) separately using images right before or after the occultation. In the other hand, we compute the 2D profile of each satellites for the given instants and obtained the simulated light flux for each satellite ($F_{S1}$, $F_{S2}$), this simulations already take into consideration the size of the satellite and the phase angle. The ratio of albedo ($A_{1}/A_{2}$) can be determined using Equation \eqref{eq:albedo}. We considered each satellite's albedo as uniform, due to the lack of information about albedo variations for the wavelength of our observations. 

\begin{equation}
    \frac{F_1}{F_2} = \frac{A_1}{A_2}.\frac{F_{s1}}{F_{s2}} ~. \label{eq:albedo}
\end{equation}

\begin{figure*}
\centering
\includegraphics[width=1.00\textwidth]{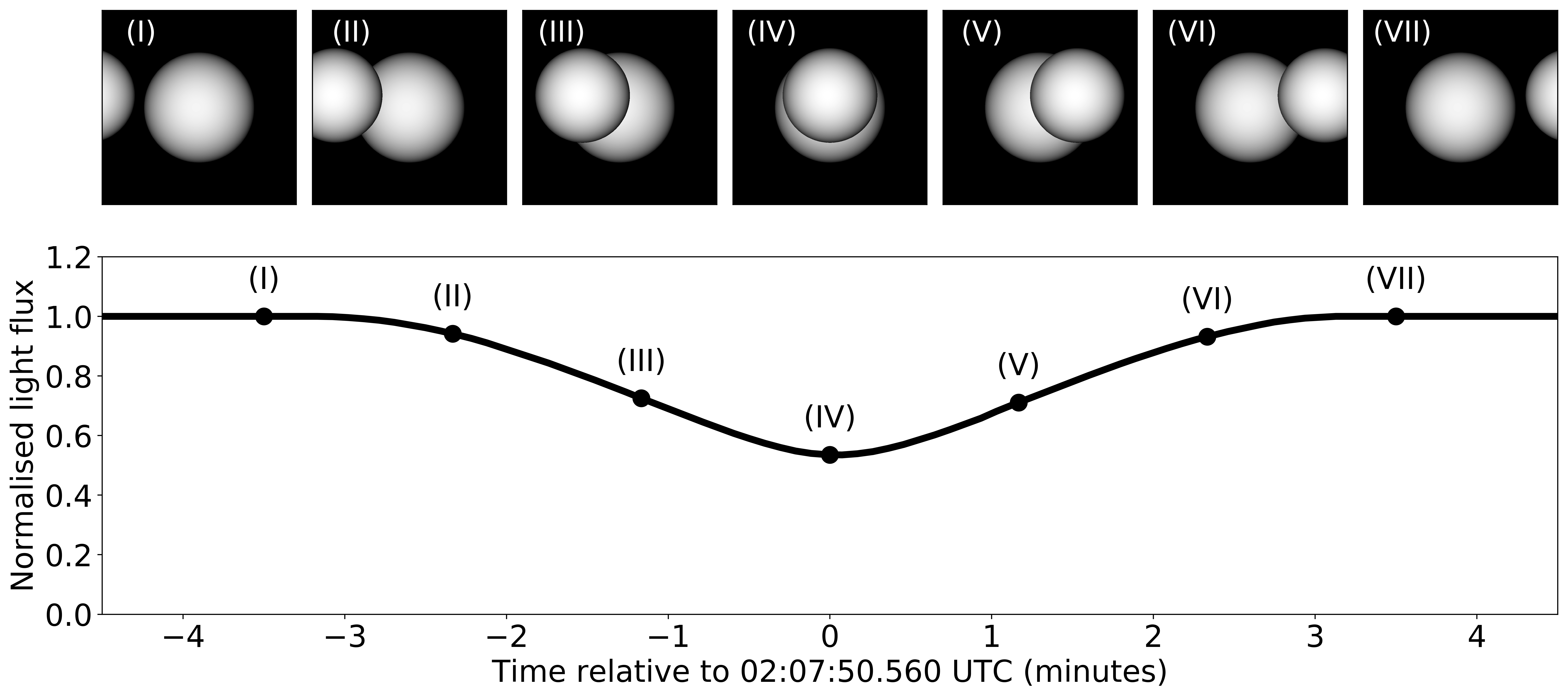}
\caption{Occultation simulation. Io was occulted by Europa on February 22, 2015, at 02:07 UTC as seen at OPD. The bottom panel shows the simulated light curve and seven instants (black dots) were highlighted. The 2D profiles for each instant are displayed in the top panel. The albedo ratio was 0.960, as determined before the event. The profile resolution is 1 mas ($\sim$ 3.2 km).} \label{Fig:occ_model}
\end{figure*}

Similar to \cite{Dias-Oliveira2013} we discretise the satellite into a 2D profile with elements of a given spatial resolution. However, unlike \cite{Dias-Oliveira2013}, we created 2D satellite profiles with much better spatial resolutions, 1 mas ($\sim$ 3 km), avoiding eventual round off errors in the simulated flux counts. The profiles were positioned for each instant of time ($t$) using the separation between the satellites, overlapping the occulted satellite with the occulting one when necessary, obtaining a new combined 2D profile of both satellites. The light flux ($F(t)$) was numerically integrated over the combined 2D profile for a given instant ($t$). Then, this is repeated for every instant ($t$) of the event to produce the simulated light curve. This simulated light curve was then normalised by the sums of individual fluxes ($F_1 + F_2$). Figure \ref{Fig:occ_model} shows a simulated light curve for the event when Europa occulted Io on February 22, 2015. The black dots indicate seven instants for which the respective 2D profiles are displayed on each corresponding box. For this event the albedo ratio was 0.960, as determined before the event.

\subsubsection{Eclipse}\label{simulation_ecl}

Following \cite{Dias-Oliveira2013}, we developed a numerical 2D mask that incorporates the two regions of the eclipsing satellite's shadow, the umbra and the penumbra. This mask was then applied to the 2D profile of the eclipsed satellite, considering the separation between both satellites as seen from the heliocentre. Once again, the spatial resolution was set as 1 mas ($\sim$ 3 km).

For the penumbra region, the fraction of sunlight that reaches the eclipsed satellite was computed by using a numerical method. The solar limb darkening was taken into account by using \cite{Hestroffer1998} empirical law, with parameters set for the 889 nm spectral region. 

\begin{figure*}
\centering
\includegraphics[width=1.00\textwidth]{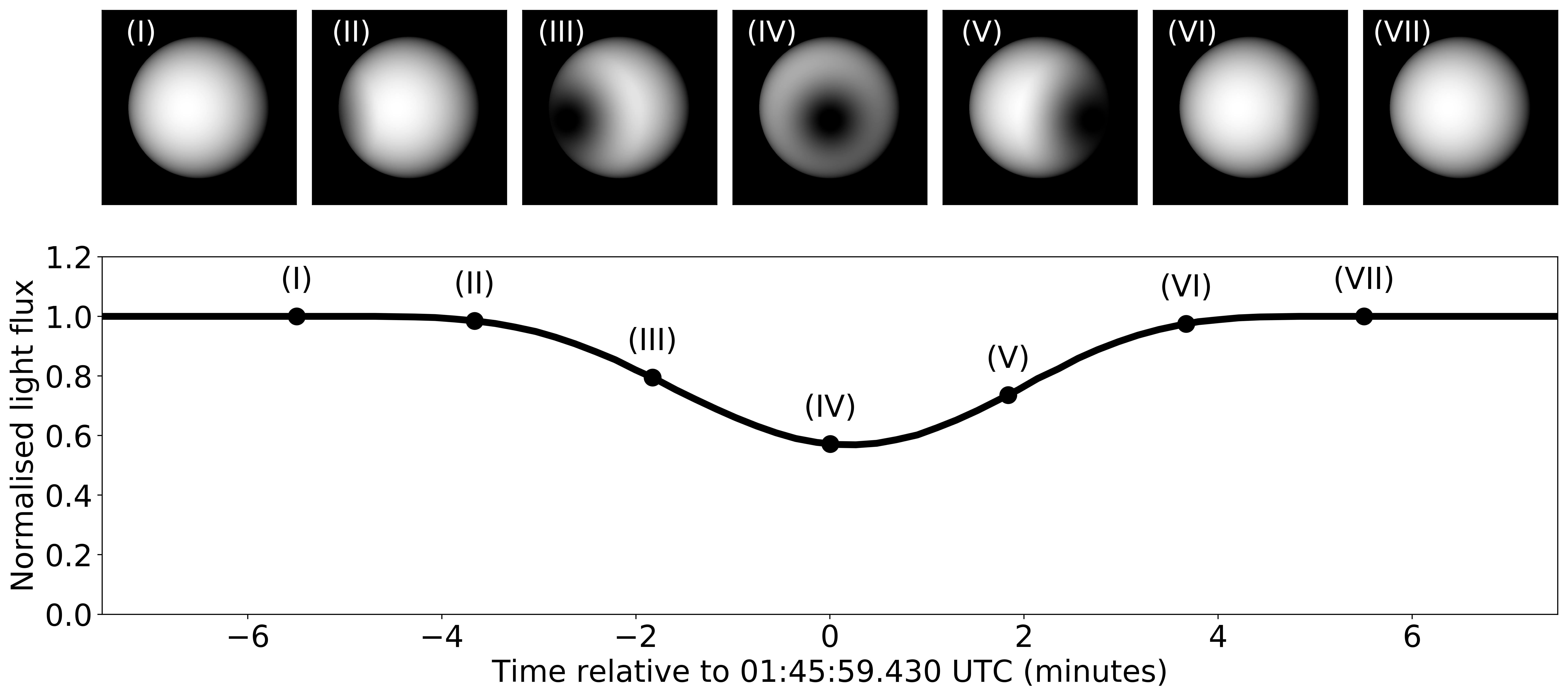}
\caption{Eclipse simulation. Ganymede was eclipsed by Europa on April 12, 2015 at 01:46 UTC as seen at FOZ. The bottom panel shows the simulated light curve and seven instants (black dots) were highlighted. The 2D profiles for each instant are displayed in the top panel. The profile resolution was 1 mas ($\sim$ 3.6 km).}
\label{Fig:ecl_model}
\end{figure*}

The light flux was numerically integrated for a given instant by using the profile after the mask was applied ($F(t)$). Then the light curve was normalised using the Light flux of the eclipsed satellite ($F_{2}$). Figure \ref{Fig:ecl_model} is a simulated light curve for the event when Europa eclipsed Ganymede on April 12, 2015. The black dots represent seven instants for which the respective 2D profiles are displayed.

\subsection{Fitting procedure}\label{fitting}

We took a somewhat different approach from that in \cite{Dias-Oliveira2013}. Here, the parameters of interest were determined by the minimisation of the Chi-square test, Eq. \eqref{eq:chi2}, where the simulated light curve is compared with the observed one. 

\begin{equation}
\chi^{2}(N-P) = \sum\dfrac{(LC_{obs} - LC_{model})^{2}}{\sigma^{2}} \label{eq:chi2}
\end{equation}

$N$ is the number of observations used in the process, and $P$ is the number of parameters fitted ($P$ = 3). $\sigma$ is the light curve's standard deviation outside the event (the noise). The parameters ($s_{0}$,$t_{0}$ and $v_{0}$) for which the chi-square is minimum ($\chi^{2}~=~\chi^{2}_{min}$) were set as the solution for the observations. The normalised $\chi^{2}$ is expected to be around 1 for good fittings.

Initially, a large range of parameter values was tested, then the ranges were narrowed as the iterative process proceeded. For computing speed, the spatial resolution was degraded in the first steps and then is set to the nominal value of 1 mas as we approached the $\chi^{2}$ minimum.

The uncertainty of each parameter ($1\sigma$ error bar) was determined by changing that parameter from its nominal solution value, so that $\chi^2$ changes from $\chi^{2}_{min}$ to $\chi^{2}_{min}~+~1$. This procedure is repeated for each parameter: $t_0$, $s_0$ and $v_0$.

The procedures described in Section \ref{redu} were developed as a \textsc{python} software that analyses and fits observed light curves. This software uses functions from \textsc{numpy}, \textsc{astropy} \citep{astropy2013}, \textsc{scipy} and \textsc{matplotlib} libraries.

\section{Results of the mutual phenomena campaigns}\label{results}

\subsection{Re-analysis of the mutual phenomena between the Galilean moons - 2009}\label{compar_2009}

From April to October of 2009, 25 light curves for 25 mutual events between the Galilean moons, 13 occultations and 12 eclipses, were observed by our group with the 60 cm Zeiss telescope of the \textit{Observat\'orio Pico dos Dias} (OPD) using the methane filter. These events were analysed by \cite{Dias-Oliveira2013} and by \cite{Arlot2014}. However, the results obtained by both presented a significant difference (higher than $3\sigma$).

More recently, \cite{Morgado2016} analysed 5 mutual approximations that were auxiliary observations designed for albedo determinations made immediately before and after mutual occultations from the 2009 campaign. The mutual approximation technique is based on the same geometrical parameters that describe a mutual occultation \citep{Morgado2016,Morgado2019}. From the measurements of the separation between both satellites, this technique allows for determining the time of maximum apparent approach between these satellites, which can be directly compared with the central instant from the occultation. The results obtained with this independent technique agree within the errors with the results obtained by \cite{Arlot2014}.

The discrepancy of the results between \cite{Dias-Oliveira2013} and \cite{Arlot2014}, the agreement between the later with \cite{Morgado2016} for 5 events, and the significant improvements in our light curve fitting procedure motivated us to re-analyse our 2009 campaign data.

Here, we present our updated results for the 2009 events in Table \ref{tb:results_2009}. Table \ref{tb:results_2009} is organised as follows: the date of the event and the satellites' pairs in the form "$S_{1} o S_{2}$" for occultation and "$S_{1} e S_{2}$" for eclipses, where 1 stands for Io, 2 for Europa, 3 for Ganymede and 4 for Callisto. We furnish the sites involved in each observation (using the alias defined in Table \ref{tb:observers}). For each event, we give the obtained central instant ($t_0$) and its uncertainty ($\sigma t_0$) in seconds of time, and the ephemeris offset ($\Delta t_0$) in mas; the impact parameter ($s_{0}$), its uncertainty ($\sigma s_0$) and its offset ($\Delta s_0$) in mas; and the relative velocity ($v_{0}$), its uncertainty ($\sigma v_0$) and its offset ($\Delta v_0$) in mas per second. All times are UTC. In the last columns, we have the rms between the observed light fluxes and the fitted ones, the number of images utilised ($N$) and the normalised $\chi^{2}$ of our fit.

\begin{table*}
\begin{center}
\caption{Updated results of the 2009 mutual phenomena campaign observed in Brazil.}
\begin{tabular}{ccccccccccccc}
\hline
\hline
Date  & Event & Obs. & $t_0$ $\pm$ $\sigma t_0$ (UTC)  & $\Delta t_0$ & $s_0$  $\pm$ $\sigma s_0$ & $\Delta s_0$ & $v_0$  $\pm$ $\sigma v_0$  & $\Delta v_0$ & rms & N & $\chi^{2}$ \\
yy-mm-dd  & $S_1$x$S_2$  &         & hh:mm:ss.s $\pm$ s.s  & mas & mas & mas & mas/s & mas/s &     \\
\hline
\hline
09-04-27 & 3o1 & OPD & 06:42:53.5 $\pm$ 0.6 & -00.8 &  121.0 $\pm$ 20.0 & -09.2 & 6.26  $\pm$  0.02 & -0.01  & 0.010 & 0801 & 1.005 \\
09-05-09 & 2o1 & OPD & 07:21:54.9 $\pm$ 0.7 & +01.5 &  538.0 $\pm$ 14.3 & -02.3 & 7.50  $\pm$  0.03 & -0.03  & 0.010 & 0580 & 1.007 \\
09-05-21 & 1o2 & OPD & 05:29:44.3 $\pm$ 0.7 & -00.2 &  033.6 $\pm$ 21.1 & +26.5 & 6.36  $\pm$  0.06 & -0.00  & 0.014 & 0869 & 1.005 \\
09-05-28 & 1o2 & OPD & 07:44:17.6 $\pm$ 0.4 & +00.0 &  219.1 $\pm$ 17.5 & +47.0 & 6.26  $\pm$  0.03 & -0.01  & 0.007 & 0801 & 1.005 \\
09-06-10 & 3e4 & OPD & 07:29:37.4 $\pm$ 1.2 & +14.2 &  315.0 $\pm$ 17.9 & +22.8 & 3.96  $\pm$  0.03 & +0.01  & 0.009 & 1285 & 1.003 \\
09-06-16 & 3e1 & OPD & 08:45:10.5 $\pm$ 1.2 & -25.6 &  928.4 $\pm$ 13.3 & +19.8 & 1.52  $\pm$  0.01 & -0.00  & 0.008 & 2697 & 1.001 \\
09-06-19 & 4e2 & OPD & 05:11:36.3 $\pm$ 0.2 & +25.1 &  489.1 $\pm$ 13.1 & +02.7 & 5.06  $\pm$  0.01 & +0.02  & 0.007 & 0901 & 1.004 \\
09-06-19 & 4e1 & OPD & 08:32:49.4 $\pm$ 1.0 & +15.6 &  930.1 $\pm$ 13.5 & +05.9 & 5.29  $\pm$  0.05 & +0.02  & 0.006 & 1201 & 1.003 \\
09-06-20 & 4e1 & OPD & 05:09:45.8 $\pm$ 1.3 & -22.6 &  530.5 $\pm$ 12.8 & +01.5 & 1.26  $\pm$  0.01 & -0.01  & 0.006 & 2326 & 1.002 \\
09-06-20 & 4e1 & OPD & 09:37:28.5 $\pm$ 1.3 & +31.6 &  412.3 $\pm$ 18.5 & -08.4 & 1.39  $\pm$  0.01 & +0.00  & 0.013 & 1704 & 1.002 \\
09-06-22 & 1o2 & OPD & 03:27:54.8 $\pm$ 1.6 & +12.0 &  576.1 $\pm$ 14.0 & +23.7 & 5.78  $\pm$  0.03 & -0.03  & 0.010 & 0905 & 1.004 \\
09-06-29 & 1o2 & OPD & 05:38:28.7 $\pm$ 0.8 & +00.5 &  606.4 $\pm$ 13.3 & +10.0 & 5.51  $\pm$  0.02 & -0.01  & 0.004 & 0801 & 1.005 \\
09-07-04 & 1e3 & OPD & 06:25:13.5 $\pm$ 0.5 & -12.7 &  395.5 $\pm$ 16.3 & -19.8 & 7.27  $\pm$  0.04 & +0.02  & 0.008 & 1641 & 1.002 \\
09-07-06 & 1e2 & OPD & 06:17:16.1 $\pm$ 1.8 & +06.2 &  718.5 $\pm$ 14.2 & -12.2 & 4.79  $\pm$  0.12 & +0.02  & 0.006 & 2001 & 1.002 \\
09-07-06 & 1o2 & OPD & 07:48:34.5 $\pm$ 0.5 & +17.4 &  603.5 $\pm$ 12.9 & -05.5 & 5.34  $\pm$  0.01 & -0.03  & 0.005 & 1004 & 1.004 \\
09-07-08 & 3e1 & OPD & 08:31:14.5 $\pm$ 0.2 & +30.1 &  223.9 $\pm$ 14.7 & -07.2 & 5.96  $\pm$  0.01 & +0.02  & 0.007 & 1758 & 1.002 \\
09-07-13 & 1e2 & OPD & 08:38:46.3 $\pm$ 1.7 & +10.3 &  623.4 $\pm$ 15.3 & -08.0 & 4.46  $\pm$  0.11 & +0.02  & 0.012 & 2001 & 1.002 \\
09-08-07 & 1e2 & OPD & 05:14:54.9 $\pm$ 1.3 & +12.4 &  444.6 $\pm$ 15.1 & +34.8 & 3.09  $\pm$  0.03 & +0.01  & 0.021 & 1775 & 1.002 \\
09-08-07 & 1o2 & OPD & 05:37:48.4 $\pm$ 0.8 & -08.8 &  283.1 $\pm$ 20.3 & -18.7 & 3.77  $\pm$  0.01 & -0.01  & 0.008 & 1664 & 1.002 \\
09-08-12 & 3o2 & OPD & 02:10:59.1 $\pm$ 3.9 & +01.0 & 1059.1 $\pm$ 13.1 & -24.7 & 2.71  $\pm$  0.02 & -0.01  & 0.004 & 1296 & 1.003 \\
09-08-22 & 1o2 & OPD & 04:07:54.9 $\pm$ 2.2 & +16.1 &  674.5 $\pm$ 12.9 & +11.2 & 1.80  $\pm$  0.01 & +0.00  & 0.004 & 2454 & 1.002 \\
09-09-16 & 1o2 & OPD & 00:46:04.4 $\pm$ 0.7 & -12.6 &  580.4 $\pm$ 15.0 & +06.5 & 3.71  $\pm$  0.02 & +0.00  & 0.011 & 0976 & 1.004 \\
09-09-16 & 1e2 & OPD & 02:15:11.0 $\pm$ 0.4 & +04.0 &  172.4 $\pm$ 12.9 & +00.8 & 3.53  $\pm$  0.02 & +0.02  & 0.007 & 1095 & 1.004 \\
09-10-24 & 3o2 & OPD & 00:35:33.9 $\pm$ 1.5 & +00.8 &  629.1 $\pm$ 13.6 & -48.8 & 4.17  $\pm$  0.02 & -0.01  & 0.008 & 1032 & 1.004 \\
09-10-25 & 1o2 & OPD & 01:21:30.8 $\pm$ 3.9 & +01.0 &  580.6 $\pm$ 17.5 & +08.5 & 5.35  $\pm$  0.07 & -0.01  & 0.014 & 0252 & 1.016 \\
\hline
\hline
\label{tb:results_2009}
\end{tabular}
\end{center}
\textit{Note}: The results for the mutual phenomena campaign of 2009. $t_0$ stand for the UTC central instant, $s_0$ is the impact parameter and $v_0$ is the apparent relative velocity in the sky plane. Also contains the uncertainty in each parameter ($\sigma t_0$, $\sigma s_0$ and $\sigma v_0$) and the difference between the fitted ones and the ones expected from the ephemeris \textit{jup310} and DE435 ($\Delta t_0$, $\Delta s_0$ and $\Delta v_0$). In the last columns, we have the rms between the observed light fluxes and the fitted ones, the number of images utilised ($N$) and the normalised $\chi^{2}$ of our fit.
\end{table*}

The corresponding inter-satellite tangential coordinates ($X$ and $Y$) in the sense occulting/eclipsing satellite minus occulted/eclipsed satellite for the central instant can be found in the supplementary material, such form is the same presented by \cite{Emelyanov2006,Emelyanov2009,Arlot2014,Saquet2018}. The plots of the re-fitted light curves are available as online material in the supplementary material. The simulated and observed \textsc{ascii} light curve files are freely available in electronic form at the NSDB\footnote{Website: \url{http://nsdb.imcce.fr/nsdb/home.html}}.

The re-analysis resulted in a mean uncertainty of 15.3 mas ($\sim$ 46 km) for the impact parameter and 4.9 mas ($\sim$ 15 km) for the central instant. In Table \ref{tb:compar_2009}, we compare the updated results with the ones from \cite{Arlot2014}, \cite{Dias-Oliveira2013} and \cite{Morgado2016}. The error of each parameter normalises the differences. If the value is less than one, both results agree within $1\sigma$. At the bottom, we have the mean difference and the standard deviation for each parameter.

\begin{table*}
\begin{center}
\caption{Comparison of the updated results for the 2009 mutual events with \cite{Arlot2014}, \cite{Dias-Oliveira2013} e \cite{Morgado2016}.}
\begin{tabular}{ccccccccc}
\hline
\hline
Date  & Event & Obs. & \multicolumn{3}{c}{Central instant ($t_0$)}  & \multicolumn{3}{c}{Impact parameter ($s_0$)}\\
yy-mm-dd      & $S_1$x$S_2$ & &  [2] - [1] & [3] - [1] & [4] - [1]  &  [2] - [1] & [3] - [1] & [4] - [1]     \\
\hline
\hline
09-04-27  & 3o1 & OPD & -0.65 &  -7.15 & --    & +0.15 & -0.74 & -- \\
09-05-09  & 2o1 & OPD & +0.78 &  -2.34 & -0.12 & +0.23 & +0.97 & -- \\
09-05-21  & 1o2 & OPD & +0.28 &  -4.94 & --    & +3.30 & +1.05 & -- \\
09-05-28  & 1o2 & OPD & +0.98 &  -9.24 & +0.57 & -0.00 & -2.36 & -- \\
09-06-10  & 3e4 & OPD & +0.13 &  +1.54 & --    & +0.13 & +1.56 & -- \\
09-06-16  & 3e1 & OPD & +0.99 &  -1.99 & --    & -0.12 & +8.81 & -- \\
09-06-19  & 4e2 & OPD & -1.87 &  +0.99 & --    & +0.03 & +4.45 & -- \\
09-06-19  & 4e1 & OPD & -0.83 &  +0.62 & --    & +0.04 & +8.26 & -- \\
09-06-20  & 4e1 & OPD & -0.70 &  +1.14 & --    & +0.08 & -4.27 & -- \\
09-06-20  & 4e1 & OPD & -0.03 &  -2.85 & --    & -0.13 & +9.77 & -- \\
09-06-22  & 1o2 & OPD & +0.27 &  -2.18 & -0.07 & +0.16 & -0.79 & -- \\
09-06-29  & 1o2 & OPD & +0.31 &  -5.30 & --    & +0.07 & -0.86 & -- \\
09-07-04  & 1e3 & OPD & -0.03 &  +1.17 & --    & -0.26 & +4.57 & -- \\
09-07-06  & 1o2 & OPD & -0.13 &  +0.25 & -0.30 & +0.14 & +5.74 & -- \\
09-07-06  & 1e2 & OPD & +0.50 &  -4.89 & --    & +0.09 & -0.01 & -- \\
09-07-08  & 3e1 & OPD & -0.61 &  +1.33 & --    & -0.04 & +1.84 & -- \\
09-07-13  & 1e2 & OPD & -0.31 &  +0.22 & --    & +0.18 & +8.36 & -- \\
09-08-07  & 1o2 & OPD & -0.22 &  +0.29 & +0.44 & +0.32 & +7.38 & -- \\
09-08-07  & 1e2 & OPD & -0.34 &  -1.71 & --    & -0.09 & -0.17 & -- \\
09-08-12  & 3o2 & OPD & +0.83 &  +0.60 & --    & -0.12 & +0.50 & -- \\
09-08-22  & 1o2 & OPD & +0.31 &  +0.29 & --    & +0.07 & -0.99 & -- \\
09-09-16  & 1e2 & OPD & -0.64 &  +3.52 & --    & -0.23 & -1.43 & -- \\
09-09-16  & 1o2 & OPD & +0.51 &  -0.43 & --    & -0.22 & +1.02 & -- \\
09-10-24  & 3o2 & OPD & -0.25 & +14.15 & --    & -0.26 & +2.78 & -- \\
09-10-25  & 1o2 & OPD & +0.62 &  +1.56 & --    & +0.23 & -0.98 & -- \\
\hline
\multicolumn{3}{c}{Mean} & -0.00 & -0.61 & +0.10 & +0.15 & +2.18 & --   \\
\multicolumn{3}{c}{Standard deviation} & 0.67 & 4.34 & 0.34 & 0.67 & 3.91  & --  \\
\hline
\hline
\label{tb:compar_2009}
\end{tabular}
\end{center}
\textit{Note}: [1] This project, [2] \cite{Arlot2014}, [3] \cite{Dias-Oliveira2013}, [4] \cite{Morgado2016}. Comparison between the different reduction process divided by the uncertainty of each parameter.
\end{table*}

The updated results now agree with those by \cite{Arlot2014} and \cite{Morgado2016} within $1\sigma$. We have a rms of 9.9 mas ($\sim$ 30 km) and 14.8 mas ($\sim$ 45 km) in comparison with the JPL's\footnote{The JPL ephemeris utilised was \textit{jup310} and DE435.} and the IMCCE's\footnote{The IMCCE ephemeris utilised was NOE-5-2010-GAL and DE435.} ephemeris.

\subsection{Mutual phenomena between the Galilean moons - 2014-2015}\label{res2015}

Here we present the results concerning the latest campaign. We obtained new 47 light curves, 31 occultations and 16 eclipses, from 40 events observed by 3 stations in the South and South-East of Brazil. 

\begin{figure}
\centering
\includegraphics[width=0.50\textwidth]{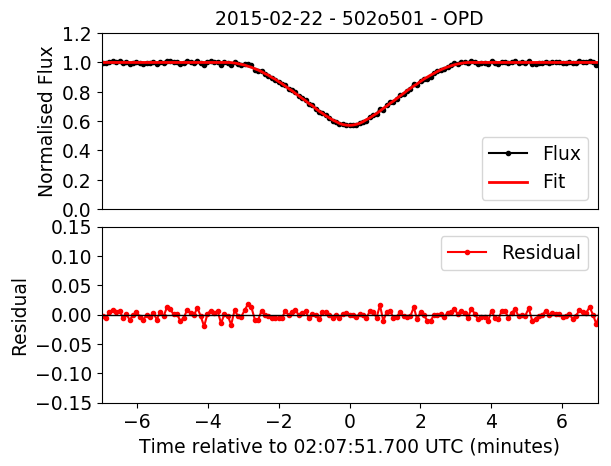}
\caption{Light curve of the event where Io was occulted by Europa on February 22 2015, observed with the 0.60 m telescope at OPD. The measured normalised flux is denoted by black dots and the fitted model represented by the red line. In the bottom panel, the red dots are the residuals of the fitting in the sense observation minus fit.}
\label{Fig:lc_occ}
\end{figure}

An example is the event where Europa occulted Io on February 22 2015. The observed light curve is illustrated in Figure \ref{Fig:lc_occ}. In the upper panel, the black dots are the light flux observed and the red line the model fitted. The bottom panel contains the residuals in the sense observation minus model. For this event, the offset for the central instant was +6.5 mas ($\sim$ 21 km) and for the impact parameter -0.8 mas ($\sim$ 3 km). The offsets regard to the JPL's \textit{jup310} and the DE435 ephemeris.

\begin{figure}
\centering
\includegraphics[width=0.50\textwidth]{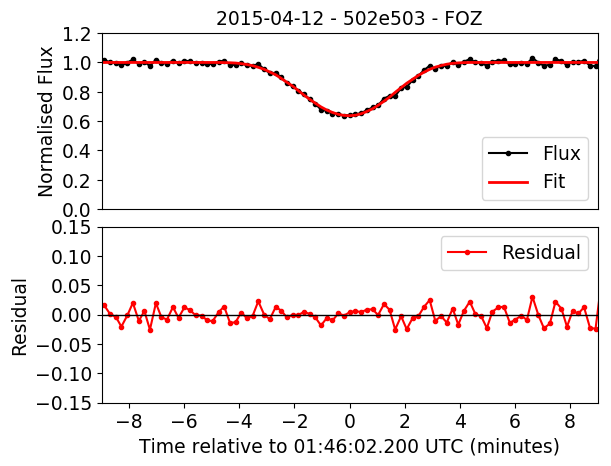}
\caption{Light curve of the event when Ganymede was eclipsed by Europa on April 12 2015, observed with the 0.28 m telescope at FOZ.}
\label{Fig:lc_ecl}
\end{figure}

A second example is the case when Europa eclipsed Ganymede on April 12 2015. The observed light curve is illustrated in Figure \ref{Fig:lc_ecl}. For this event, the offset for the central instant was +13.6 mas ($\sim$ 49 km), and for the impact parameter +9.6 mas ($\sim$ 35 km).

\begin{figure}
\centering
\includegraphics[width=0.50\textwidth]{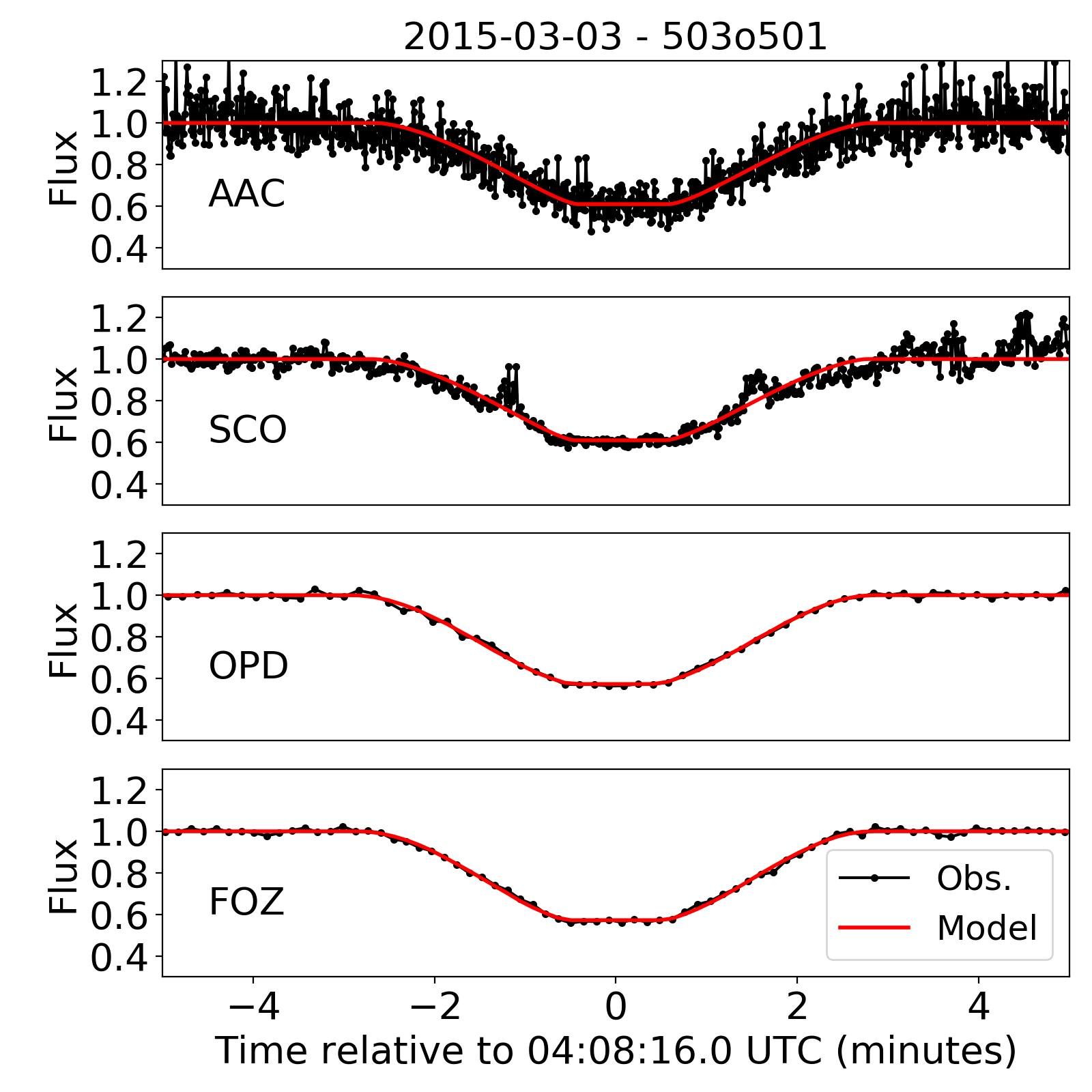}
\caption{Light curves of the event where Io was occulted by Ganymede on March 03 2015, observed at AAC, SCO, OPD and FOZ, respectively. The AAC and SCO light curves were reduced and analysed by \cite{Saquet2018}.}
\label{Fig:lc_occ_compar}
\end{figure}

The multiple coverage observational strategy reduced the number of events lost by overcast weather or instrumental issues. An example was the Io occultation by Ganymede on March 03 2015, OPD and FOZ observed this event. Also, two other stations in the USA observed this event, one in Arnold (AAC) and another in Scottsdale (SCO). These observations were made in the context of the international mutual phenomena campaign PHEMU15,  \citep{Saquet2018,Emelianov2017}. Both light curves are available at the NSDB. In the Figure \ref{Fig:lc_occ_compar} we compare our light curves (OPD and FOZ) with the ones analysed by \cite{Saquet2018} (AAC and SCO). The central instant obtained by the observations agrees within $2\sigma$. Notice that all curves present similar features and we highlight the small residual in our light curves.

\begin{table*}
\begin{center}
\caption{Results of the 2014-2015 mutual phenomena campaign observed in Brazil.}
\begin{tabular}{ccccccccccccc}
\hline
\hline
Date  & Event & Obs. & $t_0$ $\pm$ $\sigma t_0$ (UTC) & $\Delta t_0$ & $s_0$  $\pm$ $\sigma s_0$ & $\Delta s_0$ & $v_0$  $\pm$ $\sigma v_0$  & $\Delta v_0$ & rms & N & $\chi^{2}$ \\
yy-mm-dd  & $S_1$x$S_2$  &         & hh:mm:ss.s $\pm$ s.s  & mas & mas & mas & mas/s & mas/s &     \\
\hline
\hline
14-11-02 & 4o1 & OPD & 06:02:14.8 $\pm$ 1.9 & -07.2 & 288.0 $\pm$ 18.6  &  -05.0  & 1.83  $\pm$  0.03 & +0.00   & 0.041 & 230 & 1.018 \\ 
14-11-19 & 4o2 & OPD & 07:37:43.9 $\pm$ 0.2 & +00.7 & 301.3 $\pm$  6.0  &  +00.2  & 5.04  $\pm$  0.03 & +0.01   & 0.010 & 194 & 1.021 \\ 
14-12-20 & 2o1 & FOZ & 05:41:38.7 $\pm$ 0.8 & -03.0 & 162.4 $\pm$  8.5  &  -17.5  & 1.62  $\pm$  0.02 & -0.00   & 0.008 & 376 & 1.011 \\ 
14-12-21 & 4e1 & FOZ & 03:22:39.2 $\pm$ 1.5 & +00.1 & 253.3 $\pm$  9.0  &  +26.2  & 2.58  $\pm$  0.02 & +0.00   & 0.018 & 281 & 1.014 \\ 
14-12-21 & 3o1 & FOZ & 04:17:20.7 $\pm$ 3.2 & -00.8 & 455.1 $\pm$ 16.1  &  +14.0  & 3.66  $\pm$  0.03 & -0.01   & 0.151 & 244 & 1.017 \\ 
14-12-24 & 2e3 & FOZ & 06:35:06.5 $\pm$ 1.1 & -00.8 & 255.2 $\pm$  8.8  &  +09.5  & 1.93  $\pm$  0.02 & +0.00   & 0.014 & 359 & 1.011 \\ 
15-01-21 & 2e1 & FOZ & 03:52:29.3 $\pm$ 1.0 & +00.8 & 457.5 $\pm$  3.8  &  +12.8  & 3.08  $\pm$  0.04 & +0.00   & 0.008 & 141 & 1.029 \\ 
15-02-02 & 3o2 & FOZ & 07:56:42.3 $\pm$ 1.9 & -02.3 &  85.0 $\pm$ 27.4  &  +08.0  & 7.78  $\pm$  0.05 & -0.01   & 0.022 & 110 & 1.038 \\ 
15-02-22 & 2o1 & OPD & 02:07:51.7 $\pm$ 0.2 & +06.5 & 125.0 $\pm$ 12.9  &  -00.8  & 5.55  $\pm$  0.03 & -0.02   & 0.007 & 201 & 1.020 \\ 
15-02-22 & 2e1 & OPD & 02:45:11.0 $\pm$ 0.3 & +04.1 &  17.0 $\pm$  9.6  &  +07.5  & 4.62  $\pm$  0.03 & +0.02   & 0.013 & 186 & 1.022 \\ 
15-03-01 & 2o1 & FOZ & 04:12:38.3 $\pm$ 0.7 & -00.7 &  17.6 $\pm$ 10.5  &  +16.8  & 5.73  $\pm$  0.04 & -0.01   & 0.012 & 131 & 1.031 \\ 
15-03-01 & 2e1 & FOZ & 05:05:06.7 $\pm$ 2.4 & +06.6 &  94.6 $\pm$ 33.3  &  -06.8  & 4.94  $\pm$  0.04 & +0.00   & 0.059 & 132 & 1.031 \\ 
15-03-03 & 3o1 & OPD & 04:08:16.3 $\pm$ 0.5 & +00.3 &  95.0 $\pm$ 14.4  &  +32.3  & 8.53  $\pm$  0.05 & -0.03   & 0.010 & 100 & 1.042 \\ 
15-03-03 & 3o1 & FOZ & 04:08:15.5 $\pm$ 0.6 & -07.0 &  86.8 $\pm$ 10.8  &  +23.7  & 8.50  $\pm$  0.05 & -0.02   & 0.010 & 107 & 1.039 \\ 
15-03-06 & 1e2 & OPD & 01:16:16.6 $\pm$ 0.4 & +09.4 & 570.9 $\pm$  4.0  &  +06.2  & 7.50  $\pm$  0.04 & +0.04   & 0.005 & 124 & 1.033 \\ 
15-03-09 & 3e2 & OPD & 23:39:32.6 $\pm$ 0.3 & +08.1 &  67.7 $\pm$  7.2  &  +02.6  & 5.87  $\pm$  0.03 & +0.03   & 0.012 & 213 & 1.019 \\ 
15-03-13 & 1e2 & FOZ & 03:29:09.9 $\pm$ 1.1 & +10.8 & 445.2 $\pm$ 20.0  &  +03.7  & 7.42  $\pm$  0.03 & +0.03   & 0.020 & 198 & 1.021 \\ 
15-03-13 & 1e3 & FOZ & 23:29:44.5 $\pm$ 1.3 & +04.0 & 236.3 $\pm$  4.4  &  +08.5  & 1.56  $\pm$  0.02 & -0.00   & 0.015 & 408 & 1.010 \\ 
15-03-16 & 4o2 & FOZ & 01:38:58.1 $\pm$ 0.3 & -00.5 & 373.0 $\pm$  6.3  &  +02.0  & 3.59  $\pm$  0.03 & -0.00   & 0.013 & 246 & 1.017 \\ 
15-03-17 & 3e2 & FOZ & 02:53:15.2 $\pm$ 0.2 & +08.2 & 225.4 $\pm$ 14.4  &  -03.1  & 5.77  $\pm$  0.03 & +0.02   & 0.012 & 234 & 1.017 \\ 
15-03-18 & 2e1 & GOA & 22:50:43.6 $\pm$ 1.9 & +06.1 & 398.2 $\pm$ 25.2  &  -02.7  & 5.55  $\pm$  0.02 & +0.02   & 0.058 & 288 & 1.014 \\ 
15-03-24 & 3o4 & OPD & 00:14:41.4 $\pm$ 0.9 & -38.1 & 499.0 $\pm$ 13.9  &  -16.0  & 5.39  $\pm$  0.03 & -0.01   & 0.006 & 182 & 1.022 \\ 
15-03-24 & 3o4 & FOZ & 00:14:41.6 $\pm$ 0.8 & -37.5 & 519.1 $\pm$  9.9  &  +03.5  & 5.38  $\pm$  0.03 & -0.01   & 0.008 & 207 & 1.020 \\ 
15-03-25 & 2o1 & FOZ & 23:35:01.6 $\pm$ 1.1 & +00.8 & 400.2 $\pm$ 13.2  &  +01.7  & 6.34  $\pm$  0.06 & -0.02   & 0.011 & 083 & 1.051 \\ 
15-03-25 & 2o1 & GOA & 23:35:01.3 $\pm$ 0.7 & -01.5 & 398.9 $\pm$  6.6  &  +00.7  & 6.40  $\pm$  0.05 & -0.04   & 0.006 & 092 & 1.045 \\ 
15-03-26 & 2e1 & OPD & 01:07:48.0 $\pm$ 5.3 & +10.3 & 516.7 $\pm$ 44.7  &  -12.2  & 5.78  $\pm$  0.04 & +0.02   & 0.048 & 151 & 1.027 \\ 
15-04-02 & 2o1 & OPD & 01:43:55.9 $\pm$ 0.7 & +00.0 & 479.5 $\pm$  5.7  &  -00.5  & 6.46  $\pm$  0.06 & -0.02   & 0.005 & 082 & 1.051 \\ 
15-04-02 & 2o1 & FOZ & 01:43:55.6 $\pm$ 1.3 & -02.3 & 482.3 $\pm$ 15.0  &  +02.5  & 6.50  $\pm$  0.05 & -0.03   & 0.011 & 104 & 1.040 \\ 
15-04-02 & 2e1 & OPD & 03:24:16.8 $\pm$ 3.1 & +07.8 & 658.7 $\pm$ 19.1  &  -02.1  & 6.02  $\pm$  0.04 & +0.01   & 0.018 & 129 & 1.032 \\ 
15-04-03 & 1o3 & FOZ & 22:58:19.0 $\pm$ 5.9 & +05.7 & 737.3 $\pm$  6.0  &  -05.5  & 1.23  $\pm$  0.02 & +0.01   & 0.021 & 383 & 1.011 \\ 
15-04-06 & 1e2 & FOZ & 23:16:40.4 $\pm$ 0.2 & +11.3 &  54.2 $\pm$ 10.8  &  +04.2  & 6.96  $\pm$  0.04 & +0.02   & 0.007 & 152 & 1.027 \\ 
15-04-12 & 2e3 & FOZ & 01:46:02.2 $\pm$ 1.0 & +13.6 & 142.6 $\pm$  8.3  &  +09.6  & 4.96  $\pm$  0.04 & +0.00   & 0.013 & 121 & 1.034 \\ 
15-04-14 & 1e2 & OPD & 01:30:58.3 $\pm$ 0.4 & +10.4 &  49.4 $\pm$  8.5  &  -01.5  & 6.83  $\pm$  0.05 & +0.01   & 0.024 & 093 & 1.045 \\ 
15-04-17 & 4o1 & OPD & 23:47:06.9 $\pm$ 0.9 & -01.3 & 711.5 $\pm$  4.5  &  +01.8  & 5.05  $\pm$  0.05 & -0.01   & 0.007 & 101 & 1.041 \\ 
15-04-17 & 4o1 & GOA & 23:47:06.9 $\pm$ 0.9 & -01.0 & 712.1 $\pm$  5.1  &  +02.7  & 5.06  $\pm$  0.04 & -0.02   & 0.010 & 128 & 1.032 \\ 
15-04-18 & 4o3 & OPD & 01:32:30.4 $\pm$ 0.9 & -02.3 &  69.5 $\pm$ 15.1  &  +06.8  & 5.02  $\pm$  0.03 & -0.00   & 0.021 & 161 & 1.025 \\ 
15-04-18 & 1o3 & OPD & 20:54:45.6 $\pm$ 3.7 & -02.0 & 699.4 $\pm$ 33.0  &  +03.3  & 5.50  $\pm$  0.04 & +0.01   & 0.047 & 119 & 1.035 \\ 
15-04-25 & 1o3 & OPD & 23:45:28.1 $\pm$ 1.3 & +00.5 & 679.2 $\pm$  7.5  &  -02.3  & 6.05  $\pm$  0.05 & -0.02   & 0.007 & 100 & 1.042 \\ 
15-04-25 & 1o3 & FOZ & 23:45:26.7 $\pm$ 3.3 & -08.0 & 685.0 $\pm$ 28.8  &  +03.8  & 5.95  $\pm$  0.04 & +0.01   & 0.014 & 133 & 1.031 \\ 
15-04-26 & 2o1 & OPD & 21:25:00.0 $\pm$ 3.4 & +01.8 & 584.6 $\pm$ 27.6  &  +00.5  & 7.08  $\pm$  0.08 & -0.07   & 0.029 & 064 & 1.067 \\ 
15-04-29 & 3o1 & OPD & 00:29:06.9 $\pm$ 1.2 & -15.6 & 661.3 $\pm$ 19.2  &  -04.0  & 6.92  $\pm$  0.06 & -0.01   & 0.009 & 087 & 1.048 \\ 
15-04-29 & 3o1 & GOA & 00:29:07.6 $\pm$ 2.4 & -10.5 & 663.8 $\pm$ 22.2  &  -01.7  & 7.00  $\pm$  0.05 & -0.04   & 0.026 & 095 & 1.044 \\ 
15-05-03 & 2o1 & OPD & 23:39:19.6 $\pm$ 2.2 & -08.1 & 571.1 $\pm$ 25.6  &  +11.5  & 6.83  $\pm$  0.08 & -0.04   & 0.033 & 063 & 1.068 \\ 
15-05-05 & 3o2 & FOZ & 21:54:22.0 $\pm$ 2.4 & +00.3 & 780.5 $\pm$ 21.9  &  -01.7  & 5.30  $\pm$  0.04 & -0.01   & 0.010 & 117 & 1.035 \\ 
15-05-13 & 3o2 & OPD & 01:13:50.0 $\pm$ 3.0 & +06.5 & 593.2 $\pm$ 27.1  &  +03.5  & 5.04  $\pm$  0.04 & -0.01   & 0.038 & 137 & 1.030 \\ 
15-06-04 & 2o1 & FOZ & 21:55:27.9 $\pm$ 0.2 & -01.8 & 160.8 $\pm$  6.3  &  -06.8  & 6.89  $\pm$  0.05 & -0.02   & 0.009 & 094 & 1.044 \\ 
15-06-18 & 3o1 & GOA & 21:01:51.0 $\pm$ 1.1 & +01.0 & 237.6 $\pm$ 18.6  &  +10.7  & 3.82  $\pm$  0.03 & -0.02   & 0.017 & 215 & 1.019 \\ 
\hline
15-03-02 & 3e5 & 1.60 & 23:17:06.0 $\pm$ 2.3 & -22.5 & 391.3 $\pm$ 76.3 &  +20.9  & 8.44  $\pm$  0.53 & -0.09   & 0.101 & 060 & 1.080 \\ 
\hline
\hline
\label{tb:results_2014-2015}
\end{tabular}
\end{center}
\textit{Note}: Similar as the note in the Table \ref{tb:results_2009} for the mutual phenomena campaign of the 2014-2015.
\end{table*}

The results for these events are presented in Table \ref{tb:results_2014-2015}. Table \ref{tb:results_2014-2015} is organised as follows: the date of the event and the satellites' pairs in the form "$S_{1} o S_{2}$" for occultation and "$S_{1} e S_{2}$" for eclipses, where 1 stands for Io, 2 for Europa, 3 for Ganymede and 4 for Callisto. We furnish the sites involved in each observation (using the alias defined in Table \ref{tb:observers}). For each event, we give the obtained central instant ($t_0$) and its uncertainty ($\sigma t_0$) in seconds of time, and the ephemeris offset ($\Delta t_0$) in mas; the impact parameter ($s_{0}$), its uncertainty ($\sigma s_0$) and the offset ($\Delta s_0$) in mas; and the relative velocity ($v_{0}$), its uncertainty ($\sigma v_0$) and the offset ($\Delta v_0$) in mas per second. All times are UTC. In the last columns, we have the rms between the observational curves and the fitted ones, the number of images utilised ($N$) and the normalised $\chi^{2}$ of our fit.

The corresponding inter-satellite tangential coordinates ($X$ and $Y$) in the sense occulting/eclipsing satellite minus occulted/eclipsed satellite for the central instant can be found in the supplementary material, such formalism is the same presented by \cite{Emelyanov2006,Emelyanov2009,Arlot2014,Saquet2018}. The plots of the fitted light curves are available as online material in the supplementary material. The simulated and observed light curve \textsc{ascii} files are freely available in electronic form at the NSDB.

The mean uncertainty of our results is 14.8 mas ($\sim$ 45 km) for the impact parameter and 7.5 mas ($\sim$ 23 km) for the central instant. The rms relative to JPL ephemeris was 9.2 mas ($\sim$ 28 km) and 13.5 mas ($\sim$ 40 km) relative to IMCCE ephemeris. 

From the 2014-2015 events, 10 were also analysed using a different procedure and published by \cite{Saquet2018}. In average, the comparison between this procedure and ours agrees within $1\sigma$. 

\subsection{Amalthea eclipsed by Ganymede - 02 March 2015}\label{resAmalthea}

One particular event in our 2014-2015 observational campaign was the eclipse of Amalthea by Ganymede. The astrometry of this inner satellite is not easy to be done due to its proximity to Jupiter (major semi-axis equal to 2.54 Jupiter's radius). Often, coronagraphy techniques are needed to separate this object from Jupiter's scattered light \citep{Kulyk2002,Veiga2005,Robert2017}. The positional uncertainty of classical astrometry for this satellite is in the 120 mas level ($\sim$ 360 km). 

The observation of mutual eclipses involving Galilean moons and inner satellite was strongly advocated by \cite{Vachier2002}. The first registration of this kind of event was given by \cite{Christou2010}, regarding three eclipses of Amalthea observed during the 2009 mutual phenomena campaign. More recently, \cite{Saquet2016} also analysed three more eclipses of Amalthea and the first observation of an eclipse of Thebe during the 2014-2015 campaign.

Here we present the results of one eclipse of Amalthea by Ganymede observed on March, 2 of 2015 at the 1.6 m Perkin-Elmer telescope of the \textit{Observat\'orio Pico dos Dias} (OPD, MPC code: 874). This observation was made using the IKon-L CCD camera with the narrow Methane filter\footnote{Centred at 889 nm with a width of 15 nm.}.

\begin{figure*}
\centering
\includegraphics[width=1.00\textwidth]{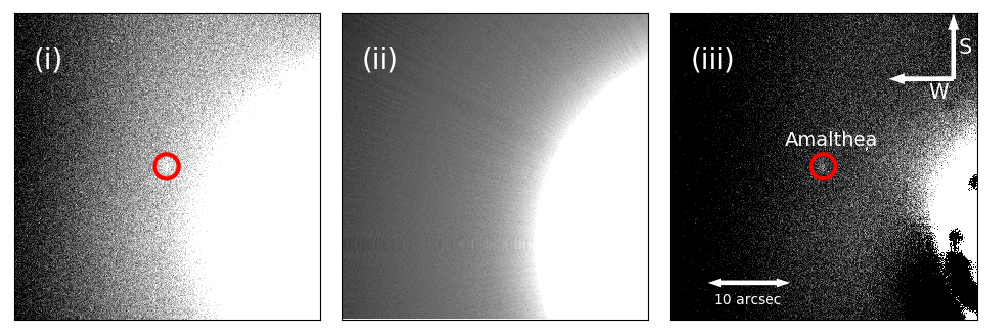}
\caption{Digital coronagraphy on an image containing part of Jupiter and its inner satellite Amalthea, as obtained on 02 March, 2015, with the 1.6 m Perkin-Elmer telescope. The left panel (i) displays the original image. The central panel (ii) shows the clean bright object profile (Jupiter) obtained. The right panel (iii) displays the final coronagraphed image (see text).}
\label{Fig:corona_amalthea}
\end{figure*}

After correcting by Bias and Flat-Field using the same procedure described above, we applied a digital coronagraphy technique to reduce the influence of Jupiter brightness in the images, this coronagraphy was done using the \textsc{praia} package \citep{Assafin2008,Assafin2009}. Briefly, the procedure is as follows. The centroid of the bright object is iteratively determined. Concentric rings with radius $R$ are formed for each image pixel at a distance $R$ to the centroid. Quartile statistics of weighted fluxes inside each ring are performed, and for each image pixel an average count is assigned. The result is an improved profile with cleaner pixel counts that better represent the bright object. The profile is then subtracted from the original image, resulting in the final coronagraphed (science) image, see Figure \ref{Fig:corona_amalthea}.

Aperture photometry was done using the \textsc{praia} package, where the size of the aperture was manually determined to maximise the signal to noise ratio. The light curve simulation and fitting procedure were the same described in Sections \ref{simulation_ecl} and \ref{fitting}. Notice that Amalthea's triaxial shape is 125 $\times$ 73 $\times$ 64 km (uncertainty of 2 km in each axis; \cite{Thomas1998}) and its rotation phase during the event was unknown. However, without any loss of precision, in our simulations, we considered Amalthea as an equivalent sphere with a radius equal to 83.5 km. Because of the time resolution of the observations (8 seconds), the spherical shape was indistinguishable from the elliptical one.

\begin{figure}
\centering
\includegraphics[width=0.50\textwidth]{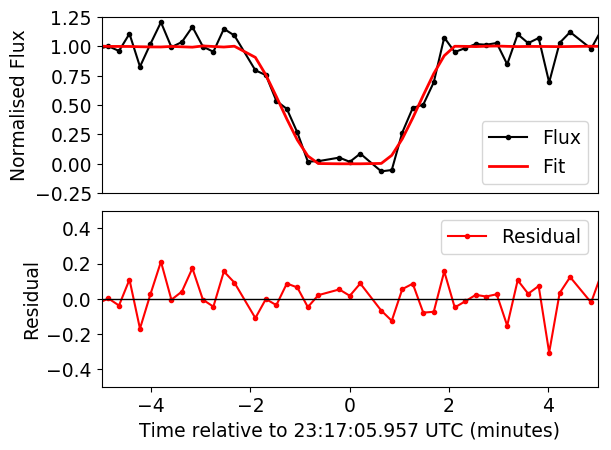}
\caption{Light curve of the event when Amalthea was eclipsed by Ganymede on March 02 2015, observed with the 1.6 m telescope at OPD.}
\label{Fig:lc_amalthea}
\end{figure}

The light curve of this event is illustrated in Fig. \ref{Fig:lc_amalthea}. We obtained a central instant with an uncertainty of 19.4 mas ($\sim$ 58.2 km) and an impact parameter with an uncertainty of 76.3 mas ($\sim$ 228 km). This corresponds to a mean uncertainty of 47.8 mas ($\sim$ 143 km). The result of this event is displayed in the last line of Table \ref{tb:results_2014-2015}. The positions obtained by \cite{Christou2010} had mean uncertainty of 82 mas ($\sim$ 246 km) and the ones obtained by \cite{Saquet2016} had a mean uncertainty of 45 mas ($\sim$ 135 km).

\section{Conclusions} \label{conclusao}

We presented in this paper the results for 40 mutual events from the observation and analysis of 47 light curves, 31 occultations and 16 eclipses, obtained during the 2014-2015 mutual phenomena campaign between the Galilean satellites. The observations were made at three stations in the South and South-East of Brazil, using telescopes with diameters ranging between 28 and 60 cm. We also obtained updated results from the re-analysis of 25 mutual events, 13 occultations and 12 eclipses, observed in Brazil by our group in 2009 with a 60 cm aperture telescope. In all observations, we used a narrow band methane filter centred at 889 nm with a width of 15 nm, that eliminates Jupiter's scattered light.

We used the Oren-Nayer model \citep{Oren1994} to characterise the reflectance of the surface of the satellites. It is a generalisation of Lambert's scattering law. The main advantage of this model is that it does not require previous knowledge about the satellite surface, and delivers excellent results. The albedo ratio was instrumentally obtained by using satellite observations before and after the mutual occultations, and the same light curve simulation routines that take into account solar phase angle and surface reflectance. In our procedure, the simulated light curves fitted to the observed ones had a normalised chi-square very close to 1.0, indicating good agreement of our model to the data.

The updated results for 2009 now agree within $1\sigma$ with those from \cite{Arlot2014} and from \cite{Morgado2016}. The internal mean uncertainty was 10.1 mas ($\sim$ 31 km). For the 2014-2015 campaign, the mean uncertainty was 11.2 mas ($\sim$ 35 km). There is no significant difference between the quality of the data for both campaigns. Our result is comparable with other independent observations during the same campaign, such as the 23 light curves observed by \cite{Vasundhara2017} and the 21 light curves observed by \cite{Zhang2019}, which report uncertainties in the 10 and 20 mas level (30 and 60 km), respectively. \cite{Saquet2018} published the light curves of the international campaign organised by the IMCCE. It reunites 609 light curves with a mean internal uncertainty of 24 mas ($\sim$ 75 km)\footnote{The standard deviation after fitting the light curves.}. From these 609 light curves, 10 were observed by our group, the parameters obtained by \cite{Saquet2018} agree, on average, within $1\sigma$ with the results presented here.

Compared to the 236 mutual events covered by the 2014-2015 international PHEMU campaign lead by the IMCCE, Observatoire de Paris \citep{Saquet2018}, the 40 ones covered by our 2014-2015 campaign (with only 5 events in common) represent a significant contribution of about 17\%. Notice that this campaign was favourable for the north hemisphere, enhancing the weight of our southern results due to parallax effects. Equally, our new results for the 25 events represent about 15\% of the 172 mutual events covered by the international PHEMU campaign of 2009 \citep{Arlot2014}. In a similar way, our result for the eclipse by Amalthea is only the 4$^{th}$ such measurement ever published for the 2014-2015 campaign, after the 3 ones observed by \cite{Saquet2016}, representing a significant contribution to the orbit of this inner satellite of Jupiter. All the data are freely available to anyone at NSDB for further research and orbital fitting.

All these results can be used to improve the orbit and ephemeris of the Galilean satellites (plus Amalthea) taking into account the tidal forces, as pointed out by \cite{Lainey2009}.    

The next mutual phenomena events for the Galilean satellites will occur in 2021 and will favour the southern hemisphere, due to Jupiter's declination. An observational campaign such as this one will be organised in due time, the prediction of these events are already in the IMCCE website\footnote{Website: \url{http://nsdb.imcce.fr/multisat/nssephme.htm}} \citep{Arlot2019}. These campaigns can increase the accuracy and precision of ephemeris and can be helpful to space missions aimed at the Jovian system. For example, we have the ESA mission JUICE\footnote{Website: \url{http://sci.esa.int/juice/}.} and NASA's mission Europa Clipper\footnote{Website: \url{https://www.nasa.gov/europa/}.}, scheduled to be launched in the next decade (2020s).

\section*{Acknowledgements}
We thank our anonymous referee and N. Emelyanov for helpful comments. This study was financed by the Coordena\c{c}\~ao de Aperfei\c{c}oamento de Pessoal de N\'ivel Superior - Brasil (CAPES) - Finance Code 001. Part of this research is suported by INCT do e-Universo, Brazil (CNPQ grants 465376/2014-2). Based in part on observations made at the Laborat\'orio Nacional de Astrof\'isica (LNA), Itajub\'a-MG, Brazil. BM thanks the CAPES/Cofecub-394/2016-05 grant. RVM acknowledges the grants: CNPq-304544/2017-5, 401903/2016-8, Capes/Cofecub-2506/2015, Faperj: PAPDRJ-45/2013 and E-26/203.026/2015. MA thanks CNPq (Grants 427700/2018-3, 310683/2017-3 and 473002/2013-2) and FAPERJ (Grant E-26/111.488/2013). JIBC acknowledges CNPq grants 308489/2013-6 and 308150/2016-3. RS and OCW acknowledges Fapesp proc. 2016/24561-0 and 2011/08171-3, CNPq proc. 312813/2013-9 and 305737/2015-5. FBR acknowledges CNPq support, proc. 309578/2017-5. GBR thanks to the support of the CAPES and FAPERJ/PAPDRJ (E26/203.173/2016) grants. ARGJ thanks FAPESP proc. 2018/11239-8. This collaboration, as part of the Encelade working group, has been supported by the International Space Sciences Institute (ISSI) in Bern, Switzerland.

\bibliographystyle{cas-model2-names}

\bibliography{Morgado_PHEMU_PSS_ref}

\begin{thebibliography}{43}
\expandafter\ifx\csname natexlab\endcsname\relax\def\natexlab#1{#1}\fi
\providecommand{\url}[1]{\texttt{#1}}
\providecommand{\href}[2]{#2}
\providecommand{\path}[1]{#1}
\providecommand{\DOIprefix}{doi:}
\providecommand{\ArXivprefix}{arXiv:}
\providecommand{\URLprefix}{URL: }
\providecommand{\Pubmedprefix}{pmid:}
\providecommand{\doi}[1]{\href{http://dx.doi.org/#1}{\path{#1}}}
\providecommand{\Pubmed}[1]{\href{pmid:#1}{\path{#1}}}
\providecommand{\bibinfo}[2]{#2}
\ifx\xfnm\relax \def\xfnm[#1]{\unskip,\space#1}\fi
\bibitem[{{Aksnes} and {Franklin}(1976)}]{Aksnes1976}
\bibinfo{author}{{Aksnes}, K.}, \bibinfo{author}{{Franklin}, F.A.},
  \bibinfo{year}{1976}.
\newblock \bibinfo{title}{{Mutual phenomena of the Galilean satellites in 1973.
  III. Final results from 91 light curves.}}
\newblock \bibinfo{journal}{Astronomical Journal} \bibinfo{volume}{81},
  \bibinfo{pages}{464--481}.
\newblock \DOIprefix\doi{10.1086/111908}.
\bibitem[{{Arlot} et~al.(2017){Arlot}, {Cooper}, {Emelyanov}, {Lainey},
  {Meunier}, {Murray}, {Oberst}, {Pascu}, {Pasewaldt}, {Robert}, {Tajeddine}
  and {Willner}}]{Arlot2017}
\bibinfo{author}{{Arlot}, J.E.}, \bibinfo{author}{{Cooper}, N.},
  \bibinfo{author}{{Emelyanov}, N.}, \bibinfo{author}{{Lainey}, V.},
  \bibinfo{author}{{Meunier}, L.E.}, \bibinfo{author}{{Murray}, C.},
  \bibinfo{author}{{Oberst}, J.}, \bibinfo{author}{{Pascu}, D.},
  \bibinfo{author}{{Pasewaldt}, A.}, \bibinfo{author}{{Robert}, V.},
  \bibinfo{author}{{Tajeddine}, R.}, \bibinfo{author}{{Willner}, K.},
  \bibinfo{year}{2017}.
\newblock \bibinfo{title}{{Natural satellites astrometric data from either
  space probes and ground-based observatories produced by the European
  consortium ``ESPaCE''}}.
\newblock \bibinfo{journal}{Notes Scientifiques et Techniques de l'Institut de
  Mecanique Celeste} \bibinfo{volume}{105}.
\bibitem[{{Arlot} and {Emelyanov}(2019)}]{Arlot2019}
\bibinfo{author}{{Arlot}, J.E.}, \bibinfo{author}{{Emelyanov}, N.},
  \bibinfo{year}{2019}.
\newblock \bibinfo{title}{{Natural satellites mutual phenomena observations:
  Achievements and future}}.
\newblock \bibinfo{journal}{Planetary and Space Science} \bibinfo{volume}{169},
  \bibinfo{pages}{70--77}.
\newblock \DOIprefix\doi{10.1016/j.pss.2019.02.004}.
\bibitem[{{Arlot} et~al.(2014a){Arlot}, {Emelyanov}, {Varfolomeev},
  {Amoss{\'e}}, {Arena}, {Assafin}, {Barbieri}, {Bolzoni}, {Bragas-Ribas},
  {Camargo}, {Casarramona}, {Casas}, {Christou}, {Colas}, {Collard}, {Combe},
  {Constantinescu}, {Dangl}, {De Cat}, {Degenhardt}, {Delcroix},
  {Dias-Oliveira}, {Dourneau}, {Douvris}, {Druon}, {Ellington}, {Estraviz},
  {Farissier}, {Farmakopoulos}, {Garlitz}, {Gault}, {George}, {Gorda},
  {Grismore}, {Guo}, {Herald}, {Ida}, {Ishida}, {Ivanov}, {Klemt}, {Koshkin},
  {Le Campion}, {Liakos}, {Liao}, {Li}, {Loader}, {Lopresti}, {Lo Savio},
  {Marchini}, {Marino}, {Masi}, {Massall{\'e}}, {Maulella}, {McFarland},
  {Miyashita}, {Napoli}, {Noyelles}, {Pauwels}, {Pavlov}, {Peng},
  {Perell{\'o}}, {Priban}, {Prost}, {Razemon}, {Rousselle}, {Rovira}, {Ruisi},
  {Ruocco}, {Salvaggio}, {Sbarufatti}, {Shakun}, {Scheck}, {Sciuto}, {da Silva
  Neto}, {Sinyaeva}, {Sofia}, {Sonka}, {Talbot}, {Tang}, {Tejfel}, {Thuillot},
  {Tigani}, {Timerson}, {Tontodonati}, {Tsamis}, {Unwin}, {Venable},
  {Vieira-Martins}, {Vilar}, {Vingerhoets}, {Watanabe}, {Yin}, {Yu} and
  {Zambelli}}]{Arlot2014}
\bibinfo{author}{{Arlot}, J.E.}, \bibinfo{author}{{Emelyanov}, N.},
  \bibinfo{author}{{Varfolomeev}, M.I.}, \bibinfo{author}{{Amoss{\'e}}, A.},
  \bibinfo{author}{{Arena}, C.}, \bibinfo{author}{{Assafin}, M.},
  \bibinfo{author}{{Barbieri}, L.}, \bibinfo{author}{{Bolzoni}, S.},
  \bibinfo{author}{{Bragas-Ribas}, F.}, \bibinfo{author}{{Camargo}, J.I.B.},
  \bibinfo{author}{{Casarramona}, F.}, \bibinfo{author}{{Casas}, R.},
  \bibinfo{author}{{Christou}, A.}, \bibinfo{author}{{Colas}, F.},
  \bibinfo{author}{{Collard}, A.}, \bibinfo{author}{{Combe}, S.},
  \bibinfo{author}{{Constantinescu}, M.}, \bibinfo{author}{{Dangl}, G.},
  \bibinfo{author}{{De Cat}, P.}, \bibinfo{author}{{Degenhardt}, S.},
  \bibinfo{author}{{Delcroix}, M.}, \bibinfo{author}{{Dias-Oliveira}, A.},
  \bibinfo{author}{{Dourneau}, G.}, \bibinfo{author}{{Douvris}, A.},
  \bibinfo{author}{{Druon}, C.}, \bibinfo{author}{{Ellington}, C.K.},
  \bibinfo{author}{{Estraviz}, G.}, \bibinfo{author}{{Farissier}, P.},
  \bibinfo{author}{{Farmakopoulos}, A.}, \bibinfo{author}{{Garlitz}, J.},
  \bibinfo{author}{{Gault}, D.}, \bibinfo{author}{{George}, T.},
  \bibinfo{author}{{Gorda}, S.Y.}, \bibinfo{author}{{Grismore}, J.},
  \bibinfo{author}{{Guo}, D.F.}, \bibinfo{author}{{Herald}, D.},
  \bibinfo{author}{{Ida}, M.}, \bibinfo{author}{{Ishida}, M.},
  \bibinfo{author}{{Ivanov}, A.V.}, \bibinfo{author}{{Klemt}, B.},
  \bibinfo{author}{{Koshkin}, N.}, \bibinfo{author}{{Le Campion}, J.F.},
  \bibinfo{author}{{Liakos}, A.}, \bibinfo{author}{{Liao}, S.L.},
  \bibinfo{author}{{Li}, S.N.}, \bibinfo{author}{{Loader}, B.},
  \bibinfo{author}{{Lopresti}, C.}, \bibinfo{author}{{Lo Savio}, E.},
  \bibinfo{author}{{Marchini}, A.}, \bibinfo{author}{{Marino}, G.},
  \bibinfo{author}{{Masi}, G.}, \bibinfo{author}{{Massall{\'e}}, A.},
  \bibinfo{author}{{Maulella}, R.}, \bibinfo{author}{{McFarland}, J.},
  \bibinfo{author}{{Miyashita}, K.}, \bibinfo{author}{{Napoli}, C.},
  \bibinfo{author}{{Noyelles}, B.}, \bibinfo{author}{{Pauwels}, T.},
  \bibinfo{author}{{Pavlov}, H.}, \bibinfo{author}{{Peng}, Q.Y.},
  \bibinfo{author}{{Perell{\'o}}, C.}, \bibinfo{author}{{Priban}, V.},
  \bibinfo{author}{{Prost}, J.}, \bibinfo{author}{{Razemon}, S.},
  \bibinfo{author}{{Rousselle}, J.P.}, \bibinfo{author}{{Rovira}, J.},
  \bibinfo{author}{{Ruisi}, R.}, \bibinfo{author}{{Ruocco}, N.},
  \bibinfo{author}{{Salvaggio}, F.}, \bibinfo{author}{{Sbarufatti}, G.},
  \bibinfo{author}{{Shakun}, L.}, \bibinfo{author}{{Scheck}, A.},
  \bibinfo{author}{{Sciuto}, C.}, \bibinfo{author}{{da Silva Neto}, D.N.},
  \bibinfo{author}{{Sinyaeva}, N.V.}, \bibinfo{author}{{Sofia}, A.},
  \bibinfo{author}{{Sonka}, A.}, \bibinfo{author}{{Talbot}, J.},
  \bibinfo{author}{{Tang}, Z.H.}, \bibinfo{author}{{Tejfel}, V.G.},
  \bibinfo{author}{{Thuillot}, W.}, \bibinfo{author}{{Tigani}, K.},
  \bibinfo{author}{{Timerson}, B.}, \bibinfo{author}{{Tontodonati}, E.},
  \bibinfo{author}{{Tsamis}, V.}, \bibinfo{author}{{Unwin}, M.},
  \bibinfo{author}{{Venable}, R.}, \bibinfo{author}{{Vieira-Martins}, R.},
  \bibinfo{author}{{Vilar}, J.}, \bibinfo{author}{{Vingerhoets}, P.},
  \bibinfo{author}{{Watanabe}, H.}, \bibinfo{author}{{Yin}, H.X.},
  \bibinfo{author}{{Yu}, Y.}, \bibinfo{author}{{Zambelli}, R.},
  \bibinfo{year}{2014}a.
\newblock \bibinfo{title}{{The PHEMU09 catalogue and astrometric results of the
  observations of the mutual occultations and eclipses of the Galilean
  satellites of Jupiter made in 2009}}.
\newblock \bibinfo{journal}{Astronomy and Astrophysics} \bibinfo{volume}{572},
  \bibinfo{pages}{A120}.
\newblock \DOIprefix\doi{10.1051/0004-6361/201423854}.
\bibitem[{{Arlot} et~al.(2014b){Arlot}, {Saquet}, {Robert} and
  {Lainey}}]{Arlot2014b}
\bibinfo{author}{{Arlot}, J.E.}, \bibinfo{author}{{Saquet}, E.},
  \bibinfo{author}{{Robert}, V.}, \bibinfo{author}{{Lainey}, V.},
  \bibinfo{year}{2014}b.
\newblock \bibinfo{title}{{The Phemu 2015 campaign of observations of the
  mutual events of the Galilean satellites of Jupiter}}, in:
  \bibinfo{booktitle}{European Planetary Science Congress}, pp.
  \bibinfo{pages}{EPSC2014--57}.
\bibitem[{{Assafin} et~al.(2008){Assafin}, {Campos}, {Vieira Martins}, {da
  Silva Neto}, {Camargo} and {Andrei}}]{Assafin2008}
\bibinfo{author}{{Assafin}, M.}, \bibinfo{author}{{Campos}, R.P.},
  \bibinfo{author}{{Vieira Martins}, R.}, \bibinfo{author}{{da Silva Neto},
  D.N.}, \bibinfo{author}{{Camargo}, J.I.B.}, \bibinfo{author}{{Andrei}, A.H.},
  \bibinfo{year}{2008}.
\newblock \bibinfo{title}{{Instrumental and digital coronagraphy for the
  observation of the Uranus satellites{\textquoteright} upcoming mutual
  events}}.
\newblock \bibinfo{journal}{Planetary and Space Science} \bibinfo{volume}{56},
  \bibinfo{pages}{1882--1887}.
\newblock \DOIprefix\doi{10.1016/j.pss.2007.05.030}.
\bibitem[{{Assafin} et~al.(2009){Assafin}, {Vieira-Martins}, {Braga-Ribas},
  {Camargo}, {da Silva Neto} and {Andrei}}]{Assafin2009}
\bibinfo{author}{{Assafin}, M.}, \bibinfo{author}{{Vieira-Martins}, R.},
  \bibinfo{author}{{Braga-Ribas}, F.}, \bibinfo{author}{{Camargo}, J.I.B.},
  \bibinfo{author}{{da Silva Neto}, D.N.}, \bibinfo{author}{{Andrei}, A.H.},
  \bibinfo{year}{2009}.
\newblock \bibinfo{title}{{Observations and Analysis of Mutual Events between
  the Uranus Main Satellites}}.
\newblock \bibinfo{journal}{Astronomical Journal} \bibinfo{volume}{137},
  \bibinfo{pages}{4046--4053}.
\newblock \DOIprefix\doi{10.1088/0004-6256/137/4/4046}.
\bibitem[{{Assafin} et~al.(2011){Assafin}, {Vieira Martins}, {Camargo},
  {Andrei}, {Da Silva Neto} and {Braga-Ribas}}]{Assafin2011}
\bibinfo{author}{{Assafin}, M.}, \bibinfo{author}{{Vieira Martins}, R.},
  \bibinfo{author}{{Camargo}, J.I.B.}, \bibinfo{author}{{Andrei}, A.H.},
  \bibinfo{author}{{Da Silva Neto}, D.N.}, \bibinfo{author}{{Braga-Ribas}, F.},
  \bibinfo{year}{2011}.
\newblock \bibinfo{title}{{PRAIA - Platform for Reduction of Astronomical
  Images Automatically}}, in: \bibinfo{booktitle}{Gaia follow-up network for
  the solar system objects : Gaia FUN-SSO workshop proceedings}, pp.
  \bibinfo{pages}{85--88}.
\bibitem[{{Astropy Collaboration} et~al.(2013){Astropy Collaboration},
  {Robitaille}, {Tollerud}, {Greenfield}, {Droettboom}, {Bray}, {Aldcroft},
  {Davis}, {Ginsburg}, {Price-Whelan}, {Kerzendorf}, {Conley}, {Crighton},
  {Barbary}, {Muna}, {Ferguson}, {Grollier}, {Parikh}, {Nair}, {Unther},
  {Deil}, {Woillez}, {Conseil}, {Kramer}, {Turner}, {Singer}, {Fox}, {Weaver},
  {Zabalza}, {Edwards}, {Azalee Bostroem}, {Burke}, {Casey}, {Crawford},
  {Dencheva}, {Ely}, {Jenness}, {Labrie}, {Lim}, {Pierfederici}, {Pontzen},
  {Ptak}, {Refsdal}, {Servillat} and {Streicher}}]{astropy2013}
\bibinfo{author}{{Astropy Collaboration}}, \bibinfo{author}{{Robitaille},
  T.P.}, \bibinfo{author}{{Tollerud}, E.J.}, \bibinfo{author}{{Greenfield},
  P.}, \bibinfo{author}{{Droettboom}, M.}, \bibinfo{author}{{Bray}, E.},
  \bibinfo{author}{{Aldcroft}, T.}, \bibinfo{author}{{Davis}, M.},
  \bibinfo{author}{{Ginsburg}, A.}, \bibinfo{author}{{Price-Whelan}, A.M.},
  \bibinfo{author}{{Kerzendorf}, W.E.}, \bibinfo{author}{{Conley}, A.},
  \bibinfo{author}{{Crighton}, N.}, \bibinfo{author}{{Barbary}, K.},
  \bibinfo{author}{{Muna}, D.}, \bibinfo{author}{{Ferguson}, H.},
  \bibinfo{author}{{Grollier}, F.}, \bibinfo{author}{{Parikh}, M.M.},
  \bibinfo{author}{{Nair}, P.H.}, \bibinfo{author}{{Unther}, H.M.},
  \bibinfo{author}{{Deil}, C.}, \bibinfo{author}{{Woillez}, J.},
  \bibinfo{author}{{Conseil}, S.}, \bibinfo{author}{{Kramer}, R.},
  \bibinfo{author}{{Turner}, J.E.H.}, \bibinfo{author}{{Singer}, L.},
  \bibinfo{author}{{Fox}, R.}, \bibinfo{author}{{Weaver}, B.A.},
  \bibinfo{author}{{Zabalza}, V.}, \bibinfo{author}{{Edwards}, Z.I.},
  \bibinfo{author}{{Azalee Bostroem}, K.}, \bibinfo{author}{{Burke}, D.J.},
  \bibinfo{author}{{Casey}, A.R.}, \bibinfo{author}{{Crawford}, S.M.},
  \bibinfo{author}{{Dencheva}, N.}, \bibinfo{author}{{Ely}, J.},
  \bibinfo{author}{{Jenness}, T.}, \bibinfo{author}{{Labrie}, K.},
  \bibinfo{author}{{Lim}, P.L.}, \bibinfo{author}{{Pierfederici}, F.},
  \bibinfo{author}{{Pontzen}, A.}, \bibinfo{author}{{Ptak}, A.},
  \bibinfo{author}{{Refsdal}, B.}, \bibinfo{author}{{Servillat}, M.},
  \bibinfo{author}{{Streicher}, O.}, \bibinfo{year}{2013}.
\newblock \bibinfo{title}{{Astropy: A community Python package for astronomy}}.
\newblock \bibinfo{journal}{Astronomy and Astrophysics} \bibinfo{volume}{558},
  \bibinfo{pages}{A33}.
\newblock \DOIprefix\doi{10.1051/0004-6361/201322068},
  \href{http://arxiv.org/abs/1307.6212}{\tt arXiv:1307.6212}.
\bibitem[{{Butcher} and {Stevens}(1981)}]{Butcher1981}
\bibinfo{author}{{Butcher}, H.}, \bibinfo{author}{{Stevens}, R.},
  \bibinfo{year}{1981}.
\newblock \bibinfo{title}{{Image Reduction and Analysis Facility Development}}.
\newblock \bibinfo{journal}{Kitt Peak National Observatory Newsletter}
  \bibinfo{volume}{16}, \bibinfo{pages}{6}.
\bibitem[{{Christou} et~al.(2010){Christou}, {Lewis}, {Roche}, {Hidas} and
  {Brown}}]{Christou2010}
\bibinfo{author}{{Christou}, A.A.}, \bibinfo{author}{{Lewis}, F.},
  \bibinfo{author}{{Roche}, P.}, \bibinfo{author}{{Hidas}, M.G.},
  \bibinfo{author}{{Brown}, T.M.}, \bibinfo{year}{2010}.
\newblock \bibinfo{title}{{Observational detection of eclipses of J5 Amalthea
  by the Galilean satellites}}.
\newblock \bibinfo{journal}{Astronomy and Astrophysics} \bibinfo{volume}{522},
  \bibinfo{pages}{A6}.
\newblock \DOIprefix\doi{10.1051/0004-6361/201014822},
  \href{http://arxiv.org/abs/1104.0042}{\tt arXiv:1104.0042}.
\bibitem[{{Dias-Oliveira} et~al.(2013){Dias-Oliveira}, {Vieira-Martins},
  {Assafin}, {Camargo}, {Braga-Ribas}, {da Silva Neto}, {Gaspar}, {Pires dos
  Santos}, {Domingos}, {Boldrin}, {Izidoro}, {Carvalho}, {Sfair}, {Sampaio} and
  {Winter}}]{Dias-Oliveira2013}
\bibinfo{author}{{Dias-Oliveira}, A.}, \bibinfo{author}{{Vieira-Martins}, R.},
  \bibinfo{author}{{Assafin}, M.}, \bibinfo{author}{{Camargo}, J.I.B.},
  \bibinfo{author}{{Braga-Ribas}, F.}, \bibinfo{author}{{da Silva Neto}, D.N.},
  \bibinfo{author}{{Gaspar}, H.S.}, \bibinfo{author}{{Pires dos Santos}, P.M.},
  \bibinfo{author}{{Domingos}, R.C.}, \bibinfo{author}{{Boldrin}, L.A.G.},
  \bibinfo{author}{{Izidoro}, A.}, \bibinfo{author}{{Carvalho}, J.P.S.},
  \bibinfo{author}{{Sfair}, R.}, \bibinfo{author}{{Sampaio}, J.C.},
  \bibinfo{author}{{Winter}, O.C.}, \bibinfo{year}{2013}.
\newblock \bibinfo{title}{{Analysis of 25 mutual eclipses and occultations
  between the Galilean satellites observed from Brazil in 2009}}.
\newblock \bibinfo{journal}{Monthly Notices of the Royal Astronomical Society}
  \bibinfo{volume}{432}, \bibinfo{pages}{225--242}.
\newblock \DOIprefix\doi{10.1093/mnras/stt447}.
\bibitem[{{Emelyanov}(2009)}]{Emelyanov2009}
\bibinfo{author}{{Emelyanov}, N.V.}, \bibinfo{year}{2009}.
\newblock \bibinfo{title}{{Mutual occultations and eclipses of the Galilean
  satellites of Jupiter in 2002-2003: final astrometric results}}.
\newblock \bibinfo{journal}{Monthly Notices of the Royal Astronomical Society}
  \bibinfo{volume}{394}, \bibinfo{pages}{1037--1044}.
\newblock \DOIprefix\doi{10.1111/j.1365-2966.2009.14398.x}.
\bibitem[{{Emel'yanov}(2017)}]{Emelianov2017}
\bibinfo{author}{{Emel'yanov}, N.V.}, \bibinfo{year}{2017}.
\newblock \bibinfo{title}{{Current problems of dynamics of moons of planets and
  binary asteroids based on observations}}.
\newblock \bibinfo{journal}{Solar System Research} \bibinfo{volume}{51},
  \bibinfo{pages}{20--37}.
\newblock \DOIprefix\doi{10.1134/S0038094617010014}.
\bibitem[{{Emelyanov} and {Gilbert}(2006)}]{Emelyanov2006}
\bibinfo{author}{{Emelyanov}, N.V.}, \bibinfo{author}{{Gilbert}, R.},
  \bibinfo{year}{2006}.
\newblock \bibinfo{title}{{Astrometric results of observations of mutual
  occultations and eclipses of the Galilean satellites of Jupiter in 2003}}.
\newblock \bibinfo{journal}{Astronomy and Astrophysics} \bibinfo{volume}{453},
  \bibinfo{pages}{1141--1149}.
\newblock \DOIprefix\doi{10.1051/0004-6361:20064810}.
\bibitem[{{Hapke}(1981)}]{Hapke1981a}
\bibinfo{author}{{Hapke}, B.}, \bibinfo{year}{1981}.
\newblock \bibinfo{title}{{Bidirectional reflectance spectroscopy. 1. Theory}}.
\newblock \bibinfo{journal}{Journal of Geophysical Research}
  \bibinfo{volume}{86}, \bibinfo{pages}{4571--4586}.
\bibitem[{{Hapke}(1984)}]{Hapke1984}
\bibinfo{author}{{Hapke}, B.}, \bibinfo{year}{1984}.
\newblock \bibinfo{title}{{Bidirectional reflectance spectroscopy 3. Correction
  for macroscopic roughness}}.
\newblock \bibinfo{journal}{Icarus} \bibinfo{volume}{59},
  \bibinfo{pages}{41--59}.
\newblock \DOIprefix\doi{10.1016/0019-1035(84)90054-X}.
\bibitem[{{Hapke}(1986)}]{Hapke1986}
\bibinfo{author}{{Hapke}, B.}, \bibinfo{year}{1986}.
\newblock \bibinfo{title}{{Bidirectional reflectance spectroscopy 4. The
  extinction coefficient and the opposition effect}}.
\newblock \bibinfo{journal}{Icarus} \bibinfo{volume}{67},
  \bibinfo{pages}{264--280}.
\newblock \DOIprefix\doi{10.1016/0019-1035(86)90108-9}.
\bibitem[{{Hapke}(2002)}]{Hapke2002}
\bibinfo{author}{{Hapke}, B.}, \bibinfo{year}{2002}.
\newblock \bibinfo{title}{{Bidirectional Reflectance Spectroscopy. 5. The
  Coherent Backscatter Opposition Effect and Anisotropic Scattering}}.
\newblock \bibinfo{journal}{Icarus} \bibinfo{volume}{157},
  \bibinfo{pages}{523--534}.
\newblock \DOIprefix\doi{10.1006/icar.2002.6853}.
\bibitem[{{Hapke}(2008)}]{Hapke2008}
\bibinfo{author}{{Hapke}, B.}, \bibinfo{year}{2008}.
\newblock \bibinfo{title}{{Bidirectional reflectance spectroscopy. 6. Effects
  of porosity}}.
\newblock \bibinfo{journal}{Icarus} \bibinfo{volume}{195},
  \bibinfo{pages}{918--926}.
\newblock \DOIprefix\doi{10.1016/j.icarus.2008.01.003}.
\bibitem[{{Hapke}(2012)}]{Hapke2012}
\bibinfo{author}{{Hapke}, B.}, \bibinfo{year}{2012}.
\newblock \bibinfo{title}{{Bidirectional reflectance spectroscopy 7. The single
  particle phase function hockey stick relation}}.
\newblock \bibinfo{journal}{Icarus} \bibinfo{volume}{221},
  \bibinfo{pages}{1079--1083}.
\newblock \DOIprefix\doi{10.1016/j.icarus.2012.10.022}.
\bibitem[{{Hapke} and {Wells}(1981)}]{Hapke1981b}
\bibinfo{author}{{Hapke}, B.}, \bibinfo{author}{{Wells}, E.},
  \bibinfo{year}{1981}.
\newblock \bibinfo{title}{{Bidirectional reflectance spectroscopy. 2.
  Experiments and observations.}}
\newblock \bibinfo{journal}{Journal of Geophysical Research}
  \bibinfo{volume}{86}, \bibinfo{pages}{3055--3060}.
\newblock \DOIprefix\doi{10.1029/JB086iB04p03055}.
\bibitem[{{Hestroffer} and {Magnan}(1998)}]{Hestroffer1998}
\bibinfo{author}{{Hestroffer}, D.}, \bibinfo{author}{{Magnan}, C.},
  \bibinfo{year}{1998}.
\newblock \bibinfo{title}{{Wavelength dependency of the Solar limb darkening}}.
\newblock \bibinfo{journal}{Astronomy and Astrophysics} \bibinfo{volume}{333},
  \bibinfo{pages}{338--342}.
\bibitem[{{Karkoschka}(1994)}]{Karkoschka1994}
\bibinfo{author}{{Karkoschka}, E.}, \bibinfo{year}{1994}.
\newblock \bibinfo{title}{{Spectrophotometry of the Jovian Planets and Titan at
  300- to 1000-nm Wavelength: The Methane Spectrum}}.
\newblock \bibinfo{journal}{Icarus} \bibinfo{volume}{111},
  \bibinfo{pages}{174--192}.
\newblock \DOIprefix\doi{10.1006/icar.1994.1139}.
\bibitem[{{Karkoschka}(1998)}]{Karkoschka1998}
\bibinfo{author}{{Karkoschka}, E.}, \bibinfo{year}{1998}.
\newblock \bibinfo{title}{{Methane, Ammonia, and Temperature Measurements of
  the Jovian Planets and Titan from CCD-Spectrophotometry}}.
\newblock \bibinfo{journal}{Icarus} \bibinfo{volume}{133},
  \bibinfo{pages}{134--146}.
\newblock \DOIprefix\doi{10.1006/icar.1998.5913}.
\bibitem[{{Kiseleva} et~al.(2008){Kiseleva}, {Kiselev}, {Kalinichenko},
  {Vasilyeva} and {Khovricheva}}]{Kiseleva2008}
\bibinfo{author}{{Kiseleva}, T.P.}, \bibinfo{author}{{Kiselev}, A.A.},
  \bibinfo{author}{{Kalinichenko}, O.A.}, \bibinfo{author}{{Vasilyeva}, N.A.},
  \bibinfo{author}{{Khovricheva}, M.L.}, \bibinfo{year}{2008}.
\newblock \bibinfo{title}{{Results of astrometric observations of Jupiter's
  Galilean satellites at the Pulkovo Observatory from 1986 to 2005}}.
\newblock \bibinfo{journal}{Solar System Research} \bibinfo{volume}{42},
  \bibinfo{pages}{414--433}.
\newblock \DOIprefix\doi{10.1134/S0038094608050055}.
\bibitem[{{Kulyk} et~al.(2002){Kulyk}, {Jockers}, {Karpov} and
  {Sergeev}}]{Kulyk2002}
\bibinfo{author}{{Kulyk}, I.}, \bibinfo{author}{{Jockers}, K.},
  \bibinfo{author}{{Karpov}, N.}, \bibinfo{author}{{Sergeev}, A.},
  \bibinfo{year}{2002}.
\newblock \bibinfo{title}{{Astrometric CCD observations of the inner Jovian
  satellites in 1999-2000}}.
\newblock \bibinfo{journal}{Astronomy and Astrophysics} \bibinfo{volume}{383},
  \bibinfo{pages}{724--728}.
\newblock \DOIprefix\doi{10.1051/0004-6361:20011770}.
\bibitem[{{Lainey} et~al.(2009){Lainey}, {Arlot}, {Karatekin} and {van
  Hoolst}}]{Lainey2009}
\bibinfo{author}{{Lainey}, V.}, \bibinfo{author}{{Arlot}, J.E.},
  \bibinfo{author}{{Karatekin}, {\"O}.}, \bibinfo{author}{{van Hoolst}, T.},
  \bibinfo{year}{2009}.
\newblock \bibinfo{title}{{Strong tidal dissipation in Io and Jupiter from
  astrometric observations}}.
\newblock \bibinfo{journal}{Nature} \bibinfo{volume}{459},
  \bibinfo{pages}{957--959}.
\newblock \DOIprefix\doi{10.1038/nature08108}.
\bibitem[{{Lainey} et~al.(2004a){Lainey}, {Arlot} and {Vienne}}]{Lainey2004b}
\bibinfo{author}{{Lainey}, V.}, \bibinfo{author}{{Arlot}, J.E.},
  \bibinfo{author}{{Vienne}, A.}, \bibinfo{year}{2004}a.
\newblock \bibinfo{title}{{New accurate ephemerides for the Galilean satellites
  of Jupiter. II. Fitting the observations}}.
\newblock \bibinfo{journal}{Astronomy and Astrophysics} \bibinfo{volume}{427},
  \bibinfo{pages}{371--376}.
\newblock \DOIprefix\doi{10.1051/0004-6361:20041271}.
\bibitem[{{Lainey} et~al.(2004b){Lainey}, {Duriez} and {Vienne}}]{Lainey2004a}
\bibinfo{author}{{Lainey}, V.}, \bibinfo{author}{{Duriez}, L.},
  \bibinfo{author}{{Vienne}, A.}, \bibinfo{year}{2004}b.
\newblock \bibinfo{title}{{New accurate ephemerides for the Galilean satellites
  of Jupiter. I. Numerical integration of elaborated equations of motion}}.
\newblock \bibinfo{journal}{Astronomy and Astrophysics} \bibinfo{volume}{420},
  \bibinfo{pages}{1171--1183}.
\newblock \DOIprefix\doi{10.1051/0004-6361:20034565}.
\bibitem[{{Lainey} et~al.(2017){Lainey}, {Jacobson}, {Tajeddine}, {Cooper},
  {Murray}, {Robert}, {Tobie}, {Guillot}, {Mathis}, {Remus}, {Desmars},
  {Arlot}, {De Cuyper}, {Dehant}, {Pascu}, {Thuillot}, {Le Poncin-Lafitte} and
  {Zahn}}]{Lainey2017}
\bibinfo{author}{{Lainey}, V.}, \bibinfo{author}{{Jacobson}, R.A.},
  \bibinfo{author}{{Tajeddine}, R.}, \bibinfo{author}{{Cooper}, N.J.},
  \bibinfo{author}{{Murray}, C.}, \bibinfo{author}{{Robert}, V.},
  \bibinfo{author}{{Tobie}, G.}, \bibinfo{author}{{Guillot}, T.},
  \bibinfo{author}{{Mathis}, S.}, \bibinfo{author}{{Remus}, F.},
  \bibinfo{author}{{Desmars}, J.}, \bibinfo{author}{{Arlot}, J.E.},
  \bibinfo{author}{{De Cuyper}, J.P.}, \bibinfo{author}{{Dehant}, V.},
  \bibinfo{author}{{Pascu}, D.}, \bibinfo{author}{{Thuillot}, W.},
  \bibinfo{author}{{Le Poncin-Lafitte}, C.}, \bibinfo{author}{{Zahn}, J.P.},
  \bibinfo{year}{2017}.
\newblock \bibinfo{title}{{New constraints on Saturn's interior from Cassini
  astrometric data}}.
\newblock \bibinfo{journal}{Icarus} \bibinfo{volume}{281},
  \bibinfo{pages}{286--296}.
\newblock \DOIprefix\doi{10.1016/j.icarus.2016.07.014},
  \href{http://arxiv.org/abs/1510.05870}{\tt arXiv:1510.05870}.
\bibitem[{{Morgado} et~al.(2016){Morgado}, {Assafin}, {Vieira-Martins},
  {Camargo}, {Dias-Oliveira} and {Gomes-J{\'u}nior}}]{Morgado2016}
\bibinfo{author}{{Morgado}, B.}, \bibinfo{author}{{Assafin}, M.},
  \bibinfo{author}{{Vieira-Martins}, R.}, \bibinfo{author}{{Camargo}, J.I.B.},
  \bibinfo{author}{{Dias-Oliveira}, A.}, \bibinfo{author}{{Gomes-J{\'u}nior},
  A.R.}, \bibinfo{year}{2016}.
\newblock \bibinfo{title}{{Astrometry of mutual approximations between natural
  satellites. Application to the Galilean moons}}.
\newblock \bibinfo{journal}{Monthly Notices of the Royal Astronomical Society}
  \bibinfo{volume}{460}, \bibinfo{pages}{4086--4097}.
\newblock \DOIprefix\doi{10.1093/mnras/stw1244},
  \href{http://arxiv.org/abs/1605.06573}{\tt arXiv:1605.06573}.
\bibitem[{{Morgado} et~al.(2019){Morgado}, {Vieira-Martins}, {Assafin},
  {Machado}, {Camargo}, {Sfair}, {Malacarne}, {Braga-Ribas}, {Robert},
  {Bassallo}, {Benedetti-Rossi}, {Boldrin}, {Borderes-Motta}, {Camargo},
  {Crispim}, {Dias-Oliveira}, {Gomes-J{\'u}nior}, {Lainey}, {Miranda}, {Moura},
  {Ribeiro}, {Santana}, {Santos-Filho}, {Trabuco}, {Winter} and
  {Yamashita}}]{Morgado2019}
\bibinfo{author}{{Morgado}, B.}, \bibinfo{author}{{Vieira-Martins}, R.},
  \bibinfo{author}{{Assafin}, M.}, \bibinfo{author}{{Machado}, D.I.},
  \bibinfo{author}{{Camargo}, J.I.B.}, \bibinfo{author}{{Sfair}, R.},
  \bibinfo{author}{{Malacarne}, M.}, \bibinfo{author}{{Braga-Ribas}, F.},
  \bibinfo{author}{{Robert}, V.}, \bibinfo{author}{{Bassallo}, T.},
  \bibinfo{author}{{Benedetti-Rossi}, G.}, \bibinfo{author}{{Boldrin}, L.A.},
  \bibinfo{author}{{Borderes-Motta}, G.}, \bibinfo{author}{{Camargo}, B.C.B.},
  \bibinfo{author}{{Crispim}, A.}, \bibinfo{author}{{Dias-Oliveira}, A.},
  \bibinfo{author}{{Gomes-J{\'u}nior}, A.R.}, \bibinfo{author}{{Lainey}, V.},
  \bibinfo{author}{{Miranda}, J.O.}, \bibinfo{author}{{Moura}, T.S.},
  \bibinfo{author}{{Ribeiro}, F.K.}, \bibinfo{author}{{Santana}, T.},
  \bibinfo{author}{{Santos-Filho}, S.}, \bibinfo{author}{{Trabuco}, L.L.},
  \bibinfo{author}{{Winter}, O.C.}, \bibinfo{author}{{Yamashita}, T.A.R.},
  \bibinfo{year}{2019}.
\newblock \bibinfo{title}{{APPROX - mutual approximations between the Galilean
  moons: the 2016-2018 observational campaign}}.
\newblock \bibinfo{journal}{Monthly Notices of the Royal Astronomical Society}
  \bibinfo{volume}{482}, \bibinfo{pages}{5190--5200}.
\newblock \DOIprefix\doi{10.1093/mnras/sty3040},
  \href{http://arxiv.org/abs/1811.02913}{\tt arXiv:1811.02913}.
\bibitem[{{Oren} and {Nayar}(1994)}]{Oren1994}
\bibinfo{author}{{Oren}, M.}, \bibinfo{author}{{Nayar}, S.K.},
  \bibinfo{year}{1994}.
\newblock \bibinfo{title}{Generalization of lambert's reflectance model}, in:
  \bibinfo{booktitle}{Proceedings of the 21st Annual Conference on Computer
  Graphics and Interactive Techniques}, \bibinfo{publisher}{ACM},
  \bibinfo{address}{New York, NY, USA}. pp. \bibinfo{pages}{239--246}.
\newblock \URLprefix \url{http://doi.acm.org/10.1145/192161.192213},
  \DOIprefix\doi{10.1145/192161.192213}.
\bibitem[{{Peng} et~al.(2012){Peng}, {He}, {Lainey} and {Vienne}}]{Peng2012}
\bibinfo{author}{{Peng}, Q.Y.}, \bibinfo{author}{{He}, H.F.},
  \bibinfo{author}{{Lainey}, V.}, \bibinfo{author}{{Vienne}, A.},
  \bibinfo{year}{2012}.
\newblock \bibinfo{title}{{Precise CCD positions of Galilean satellite-pairs}}.
\newblock \bibinfo{journal}{Monthly Notices of the Royal Astronomical Society}
  \bibinfo{volume}{419}, \bibinfo{pages}{1977--1982}.
\newblock \DOIprefix\doi{10.1111/j.1365-2966.2011.19852.x}.
\bibitem[{{Robert} et~al.(2017){Robert}, {Saquet}, {Colas} and
  {Arlot}}]{Robert2017}
\bibinfo{author}{{Robert}, V.}, \bibinfo{author}{{Saquet}, E.},
  \bibinfo{author}{{Colas}, F.}, \bibinfo{author}{{Arlot}, J.E.},
  \bibinfo{year}{2017}.
\newblock \bibinfo{title}{{CCD astrometric observations of Amalthea and Thebe
  in the Gaia era}}.
\newblock \bibinfo{journal}{Monthly Notices of the Royal Astronomical Society}
  \bibinfo{volume}{467}, \bibinfo{pages}{694--698}.
\newblock \DOIprefix\doi{10.1093/mnras/stx123}.
\bibitem[{{Saquet} et~al.(2016){Saquet}, {Emelyanov}, {Colas}, {Arlot},
  {Robert}, {Christophe} and {Dechambre}}]{Saquet2016}
\bibinfo{author}{{Saquet}, E.}, \bibinfo{author}{{Emelyanov}, N.},
  \bibinfo{author}{{Colas}, F.}, \bibinfo{author}{{Arlot}, J.E.},
  \bibinfo{author}{{Robert}, V.}, \bibinfo{author}{{Christophe}, B.},
  \bibinfo{author}{{Dechambre}, O.}, \bibinfo{year}{2016}.
\newblock \bibinfo{title}{{Eclipses of the inner satellites of Jupiter observed
  in 2015}}.
\newblock \bibinfo{journal}{Astronomy and Astrophysics} \bibinfo{volume}{591},
  \bibinfo{pages}{A42}.
\newblock \DOIprefix\doi{10.1051/0004-6361/201628246},
  \href{http://arxiv.org/abs/1605.06935}{\tt arXiv:1605.06935}.
\bibitem[{{Saquet} et~al.(2018){Saquet}, {Emelyanov}, {Robert}, {Arlot},
  {Anbazhagan}, {Bailli{\'e}}, {Bardecker}, {Berezhnoy}, {Bretton}, {Campos},
  {Capannoli}, {Carry}, {Castet}, {Charbonnier}, {Chernikov}, {Christou},
  {Colas}, {Coliac}, {Dangl}, {Dechambre}, {Delcroix}, {Dias-Oliveira},
  {Drillaud}, {Duchemin}, {Dunford}, {Dupouy}, {Ellington}, {Fabre},
  {Filippov}, {Finnegan}, {Foglia}, {Font}, {Gaillard}, {Galli}, {Garlitz},
  {Gasmi}, {Gaspar}, {Gault}, {Gazeas}, {George}, {Gorda}, {Gorshanov},
  {Gualdoni}, {Guhl}, {Halir}, {Hanna}, {Henry}, {Herald}, {Houdin}, {Ito},
  {Izmailov}, {Jacobsen}, {Jones}, {Kamoun}, {Kardasis}, {Karimov},
  {Khovritchev}, {Kulikova}, {Laborde}, {Lainey}, {Lavayssiere}, {Le Guen},
  {Leroy}, {Loader}, {Lopez}, {Lyashenko}, {Lyssenko}, {Machado}, {Maigurova},
  {Manek}, {Marchini}, {Midavaine}, {Montier}, {Morgado}, {Naumov}, {Nedelcu},
  {Newman}, {Ohlert}, {Oksanen}, {Pavlov}, {Petrescu}, {Pomazan}, {Popescu},
  {Pratt}, {Raskhozhev}, {Resch}, {Robilliard}, {Roschina}, {Rothenberg},
  {Rottenborn}, {Rusov}, {Saby}, {Saya}, {Selvakumar}, {Signoret},
  {Slesarenko}, {Sokov}, {Soldateschi}, {Sonka}, {Soulie}, {Talbot}, {Tejfel},
  {Thuillot}, {Timerson}, {Toma}, {Torsellini}, {Trabuco}, {Traverse},
  {Tsamis}, {Unwin}, {Abbeel}, {Vand enbruaene}, {Vasundhara}, {Velikodsky},
  {Vienne}, {Vilar}, {Vugnon}, {Wuensche} and {Zeleny}}]{Saquet2018}
\bibinfo{author}{{Saquet}, E.}, \bibinfo{author}{{Emelyanov}, N.},
  \bibinfo{author}{{Robert}, V.}, \bibinfo{author}{{Arlot}, J.E.},
  \bibinfo{author}{{Anbazhagan}, P.}, \bibinfo{author}{{Bailli{\'e}}, K.},
  \bibinfo{author}{{Bardecker}, J.}, \bibinfo{author}{{Berezhnoy}, A.A.},
  \bibinfo{author}{{Bretton}, M.}, \bibinfo{author}{{Campos}, F.},
  \bibinfo{author}{{Capannoli}, L.}, \bibinfo{author}{{Carry}, B.},
  \bibinfo{author}{{Castet}, M.}, \bibinfo{author}{{Charbonnier}, Y.},
  \bibinfo{author}{{Chernikov}, M.M.}, \bibinfo{author}{{Christou}, A.},
  \bibinfo{author}{{Colas}, F.}, \bibinfo{author}{{Coliac}, J.F.},
  \bibinfo{author}{{Dangl}, G.}, \bibinfo{author}{{Dechambre}, O.},
  \bibinfo{author}{{Delcroix}, M.}, \bibinfo{author}{{Dias-Oliveira}, A.},
  \bibinfo{author}{{Drillaud}, C.}, \bibinfo{author}{{Duchemin}, Y.},
  \bibinfo{author}{{Dunford}, R.}, \bibinfo{author}{{Dupouy}, P.},
  \bibinfo{author}{{Ellington}, C.}, \bibinfo{author}{{Fabre}, P.},
  \bibinfo{author}{{Filippov}, V.A.}, \bibinfo{author}{{Finnegan}, J.},
  \bibinfo{author}{{Foglia}, S.}, \bibinfo{author}{{Font}, D.},
  \bibinfo{author}{{Gaillard}, B.}, \bibinfo{author}{{Galli}, G.},
  \bibinfo{author}{{Garlitz}, J.}, \bibinfo{author}{{Gasmi}, A.},
  \bibinfo{author}{{Gaspar}, H.S.}, \bibinfo{author}{{Gault}, D.},
  \bibinfo{author}{{Gazeas}, K.}, \bibinfo{author}{{George}, T.},
  \bibinfo{author}{{Gorda}, S.Y.}, \bibinfo{author}{{Gorshanov}, D.L.},
  \bibinfo{author}{{Gualdoni}, C.}, \bibinfo{author}{{Guhl}, K.},
  \bibinfo{author}{{Halir}, K.}, \bibinfo{author}{{Hanna}, W.},
  \bibinfo{author}{{Henry}, X.}, \bibinfo{author}{{Herald}, D.},
  \bibinfo{author}{{Houdin}, G.}, \bibinfo{author}{{Ito}, Y.},
  \bibinfo{author}{{Izmailov}, I.S.}, \bibinfo{author}{{Jacobsen}, J.},
  \bibinfo{author}{{Jones}, A.}, \bibinfo{author}{{Kamoun}, S.},
  \bibinfo{author}{{Kardasis}, E.}, \bibinfo{author}{{Karimov}, A.M.},
  \bibinfo{author}{{Khovritchev}, M.Y.}, \bibinfo{author}{{Kulikova}, A.M.},
  \bibinfo{author}{{Laborde}, J.}, \bibinfo{author}{{Lainey}, V.},
  \bibinfo{author}{{Lavayssiere}, M.}, \bibinfo{author}{{Le Guen}, P.},
  \bibinfo{author}{{Leroy}, A.}, \bibinfo{author}{{Loader}, B.},
  \bibinfo{author}{{Lopez}, O.C.}, \bibinfo{author}{{Lyashenko}, A.Y.},
  \bibinfo{author}{{Lyssenko}, P.G.}, \bibinfo{author}{{Machado}, D.I.},
  \bibinfo{author}{{Maigurova}, N.}, \bibinfo{author}{{Manek}, J.},
  \bibinfo{author}{{Marchini}, A.}, \bibinfo{author}{{Midavaine}, T.},
  \bibinfo{author}{{Montier}, J.}, \bibinfo{author}{{Morgado}, B.E.},
  \bibinfo{author}{{Naumov}, K.N.}, \bibinfo{author}{{Nedelcu}, A.},
  \bibinfo{author}{{Newman}, J.}, \bibinfo{author}{{Ohlert}, J.M.},
  \bibinfo{author}{{Oksanen}, A.}, \bibinfo{author}{{Pavlov}, H.},
  \bibinfo{author}{{Petrescu}, E.}, \bibinfo{author}{{Pomazan}, A.},
  \bibinfo{author}{{Popescu}, M.}, \bibinfo{author}{{Pratt}, A.},
  \bibinfo{author}{{Raskhozhev}, V.N.}, \bibinfo{author}{{Resch}, J.M.},
  \bibinfo{author}{{Robilliard}, D.}, \bibinfo{author}{{Roschina}, E.},
  \bibinfo{author}{{Rothenberg}, E.}, \bibinfo{author}{{Rottenborn}, M.},
  \bibinfo{author}{{Rusov}, S.A.}, \bibinfo{author}{{Saby}, F.},
  \bibinfo{author}{{Saya}, L.F.}, \bibinfo{author}{{Selvakumar}, G.},
  \bibinfo{author}{{Signoret}, F.}, \bibinfo{author}{{Slesarenko}, V.Y.},
  \bibinfo{author}{{Sokov}, E.N.}, \bibinfo{author}{{Soldateschi}, J.},
  \bibinfo{author}{{Sonka}, A.}, \bibinfo{author}{{Soulie}, G.},
  \bibinfo{author}{{Talbot}, J.}, \bibinfo{author}{{Tejfel}, V.G.},
  \bibinfo{author}{{Thuillot}, W.}, \bibinfo{author}{{Timerson}, B.},
  \bibinfo{author}{{Toma}, R.}, \bibinfo{author}{{Torsellini}, S.},
  \bibinfo{author}{{Trabuco}, L.L.}, \bibinfo{author}{{Traverse}, P.},
  \bibinfo{author}{{Tsamis}, V.}, \bibinfo{author}{{Unwin}, M.},
  \bibinfo{author}{{Abbeel}, F.V.D.}, \bibinfo{author}{{Vand enbruaene}, H.},
  \bibinfo{author}{{Vasundhara}, R.}, \bibinfo{author}{{Velikodsky}, Y.I.},
  \bibinfo{author}{{Vienne}, A.}, \bibinfo{author}{{Vilar}, J.},
  \bibinfo{author}{{Vugnon}, J.M.}, \bibinfo{author}{{Wuensche}, N.},
  \bibinfo{author}{{Zeleny}, P.}, \bibinfo{year}{2018}.
\newblock \bibinfo{title}{{The PHEMU15 catalogue and astrometric results of the
  Jupiter's Galilean satellite mutual occultation and eclipse observations made
  in 2014-2015}}.
\newblock \bibinfo{journal}{Monthly Notices of the Royal Astronomical Society}
  \bibinfo{volume}{474}, \bibinfo{pages}{4730--4739}.
\newblock \DOIprefix\doi{10.1093/mnras/stx2957}.
\bibitem[{{Thomas} et~al.(1998){Thomas}, {Burns}, {Rossier}, {Simonelli},
  {Veverka}, {Chapman}, {Klaasen}, {Johnson}, {Belton} and {Galileo Solid State
  Imaging Team}}]{Thomas1998}
\bibinfo{author}{{Thomas}, P.C.}, \bibinfo{author}{{Burns}, J.A.},
  \bibinfo{author}{{Rossier}, L.}, \bibinfo{author}{{Simonelli}, D.},
  \bibinfo{author}{{Veverka}, J.}, \bibinfo{author}{{Chapman}, C.R.},
  \bibinfo{author}{{Klaasen}, K.}, \bibinfo{author}{{Johnson}, T.V.},
  \bibinfo{author}{{Belton}, M.J.S.}, \bibinfo{author}{{Galileo Solid State
  Imaging Team}}, \bibinfo{year}{1998}.
\newblock \bibinfo{title}{{The Small Inner Satellites of Jupiter}}.
\newblock \bibinfo{journal}{Icarus} \bibinfo{volume}{135},
  \bibinfo{pages}{360--371}.
\newblock \DOIprefix\doi{10.1006/icar.1998.5976}.
\bibitem[{{Vachier} et~al.(2002){Vachier}, {Arlot} and
  {Thuillot}}]{Vachier2002}
\bibinfo{author}{{Vachier}, F.}, \bibinfo{author}{{Arlot}, J.E.},
  \bibinfo{author}{{Thuillot}, W.}, \bibinfo{year}{2002}.
\newblock \bibinfo{title}{{Mutual phenomena involving J5 Amalthea in
  2002-2003}}.
\newblock \bibinfo{journal}{Astronomy and Astrophysics} \bibinfo{volume}{394},
  \bibinfo{pages}{L19--L21}.
\newblock \DOIprefix\doi{10.1051/0004-6361:20021329}.
\bibitem[{{Vasundhara} et~al.(2017){Vasundhara}, {Selvakumar} and
  {Anbazhagan}}]{Vasundhara2017}
\bibinfo{author}{{Vasundhara}, R.}, \bibinfo{author}{{Selvakumar}, G.},
  \bibinfo{author}{{Anbazhagan}, P.}, \bibinfo{year}{2017}.
\newblock \bibinfo{title}{{Analysis of mutual events of Galilean satellites
  observed from VBO during 2014-2015}}.
\newblock \bibinfo{journal}{Monthly Notices of the Royal Astronomical Society}
  \bibinfo{volume}{468}, \bibinfo{pages}{501--508}.
\newblock \DOIprefix\doi{10.1093/mnras/stx437},
  \href{http://arxiv.org/abs/1704.03518}{\tt arXiv:1704.03518}.
\bibitem[{{Veiga} and {Vieira Martins}(2005)}]{Veiga2005}
\bibinfo{author}{{Veiga}, C.H.}, \bibinfo{author}{{Vieira Martins}, R.},
  \bibinfo{year}{2005}.
\newblock \bibinfo{title}{{CCD astrometric observations of Amalthea and
  Thebe}}.
\newblock \bibinfo{journal}{Astronomy and Astrophysics} \bibinfo{volume}{437},
  \bibinfo{pages}{1147--1150}.
\newblock \DOIprefix\doi{10.1051/0004-6361:20042387}.
\bibitem[{{Zhang} et~al.(2019){Zhang}, {Han} and {Arlot}}]{Zhang2019}
\bibinfo{author}{{Zhang}, X.L.}, \bibinfo{author}{{Han}, X.L.},
  \bibinfo{author}{{Arlot}, J.E.}, \bibinfo{year}{2019}.
\newblock \bibinfo{title}{{Mutual events between Galilean satellites observed
  with SARA 0.9 m and 0.6 m telescopes during 2014-2015}}.
\newblock \bibinfo{journal}{Monthly Notices of the Royal Astronomical Society}
  \bibinfo{volume}{483}, \bibinfo{pages}{4518--4524}.
\newblock \DOIprefix\doi{10.1093/mnras/sty3030}.

\end{thebibliography}


\end{document}